\documentclass[fleqn,usenatbib]{mnras}

\usepackage{newtxtext,newtxmath}

\usepackage[T1]{fontenc}

\DeclareRobustCommand{\VAN}[3]{#2}
\let\VANthebibliography\thebibliography
\def\thebibliography{\DeclareRobustCommand{\VAN}[3]{##3}\VANthebibliography}

\usepackage{xcolor}

\usepackage{graphicx}	
\usepackage{amsmath}	

\newcommand{\hmol}{\ensuremath{\mathrm{H}_2}}
\newcommand{\hii}{H\,\textsc{ii}}

\newcommand{\oii}[1]{[\ion{O}{ii}]#1}
\newcommand{\oiii}[1]{[\ion{O}{iii}]#1}
\newcommand{\nii}[1]{[\ion{N}{ii}]#1}
\newcommand{\sii}[1]{[\ion{S}{ii}]#1}
\newcommand{\feii}[1]{[\ion{Fe}{ii}]#1}
\newcommand{\neiv}[1]{[\textrm{Ne}\textsc{\,iv}]#1}
\newcommand{\neiii}[1]{[\textrm{Ne}\textsc{\,iii}]#1}
\newcommand{\nev}[1]{[\textrm{Ne}\textsc{\,v}]#1}
\newcommand{\hei}[1]{He\textsc{\,i} #1}
\newcommand{\heii}[1]{He\textsc{\,ii} #1}
\newcommand{\siii}[1]{[\textsc{S\,iii}]#1}
\newcommand{\ciii}[1]{\textsc{C\,iii}]#1}
\newcommand{\cii}[1]{\textsc{C\,ii}]#1}
\newcommand{\ha}{\ensuremath{\mathrm{H}\alpha}}
\newcommand{\hb}{\ensuremath{\mathrm{H}\beta}}
\newcommand{\hg}{\ensuremath{\mathrm{H}\gamma}}
\newcommand{\hd}{\ensuremath{\mathrm{H}\delta}}
\newcommand{\Pa}{\ensuremath{\mathrm{Pa}\alpha}}
\newcommand{\Pb}{\ensuremath{\mathrm{Pa}\beta}}
\newcommand{\Pg}{\ensuremath{\mathrm{Pa}\gamma}}
\newcommand{\Pd}{\ensuremath{\mathrm{Pa}\delta}}
\newcommand{\Brd}{\ensuremath{\mathrm{Br}\delta}}
\newcommand{\Brg}{\ensuremath{\mathrm{Br}\gamma}}

\newcommand{\mum}{\ensuremath{\mu\mathrm{m}}}

\newcommand{\new}[1]{{ #1}}

\title[JWST and analogues]{High-z galaxies with JWST and local
  analogues --- it is not only star formation}

\author[J. Brinchmann]{
Jarle Brinchmann,$^{1}$\thanks{E-mail: jarle@astro.up.pt}
\\
$^{1}$Instituto de Astrof{\'\i}sica e Ci{\^e}ncias do Espaço,
Universidade do Porto, CAUP, Rua das Estrelas, PT4150-762 Porto,
Portugal 
}

\date{Accepted XXX. Received YYY; in original form ZZZ}

\pubyear{2022}

\begin{document}
\label{firstpage}
\pagerange{\pageref{firstpage}--\pageref{lastpage}}
\maketitle

\begin{abstract}
  I present an analysis of the JWST NIRSpec data of SMACS 0723
  released as Early Release Observations. As part of this three new
  redshifts are provided, bringing the total of reliable redshifts to
  14. I propose a modification to the direct abundance determination
  method that reduces sensitivity to flux calibration uncertainties by
  a factor of $\sim 3$ and show that the resulting abundances are in
  good agreement with Bayesian photoionization models of the
  rest-frame optical spectrum. I also show that 6355 is most likely a
  narrow-line active galactic nucleus (AGN) with
  $M_*<10^9\,\mathrm{M}_\odot$ at $z=7.66$, and argue that 10612 might
  also have an AGN contribution to its flux through comparison to
  photoionization models and low-redshift analogues. Under the
  assumption that the lines come from star-formation I find that the
  galaxies have gas depletion times of $\sim 10^7$ years, comparable
  to similar galaxies locally. I also identify a population of
  possibly shock-dominated galaxies at $z<3$ whose near-IR emission
  lines plausibly come nearly all from shocks and discuss their
  implications.  I close with a discussion of the potential for biases
  in the determination of the mass-metallicity relation using samples
  defined by detected \oiii{4363} and show using low-$z$ galaxies that
  this can lead to biases of up to 0.5 dex with a systematic trend
  with mass.
\end{abstract}


\section{Introduction}

The study of the ionized gas in galaxies has a long and storied
history with much work in particular focused on low-redshift
galaxies. One strand in these investigations has been the study of
low-redshift galaxies as analogues to high redshift galaxies. This has
been particularly focused on studies of low-metallicity star-forming
galaxies
\citep[e.g.][]{izotovPrimordialHeliumAbundance1994,kunthMostMetalpoorGalaxies2000a,papaderosExtremelyMetalpoorStarforming2008,papaderosZw18Morphological2012}
as they are thought to offer the most promising sites to study star
formation under conditions similar to the high-redshift Universe. In
parallel to this, the advent of large spectroscopic catalogues of
galaxies through, in particular, the Sloan Digital Sky Survey
\citep{2000AJ....120.1579Y} saw systematic searches for extreme
emission line galaxies that have been considered to provide examples
of galaxies where star formation conditions are similar to those at
high redshift
\citep[e.g.][]{2006A&A...448..955I,2008A&A...485..657B,cardamoneGalaxyZooGreen2009a,perez-monteroExtremeEmissionlineGalaxies2021a}. As
fashion has taken the field, the foucs has been on \hii\ galaxies
\citep[e.g.][]{tellesMorphologyIIGalaxies1997}, Blue Compact Dwarfs
\citep[BCDs, ][]{kunthMostMetalpoorGalaxies2000a}, Lyman-break
analogues
\citep{overzierLocalLymanBreak2009,heckmanExtremeFeedbackEpoch2011a,wuStarformingInterstellarMedium2019},
peas of different colours
\citep{cardamoneGalaxyZooGreen2009a,yangLyaProfileDust2017,brunkerPropertiesKISSGreen2020,yangBlueberryGalaxiesLowest2017},
and extreme line emitters pushed out to ever higher redshift
\citep[e.g.][]{2012MNRAS.421.1043S,bianLocalAnalogsHighredshift2016,masedaNatureExtremeEmission2014,amorinExtremeEmissionlineGalaxies2015}. There
are significant overlaps between most of these classes and they are
typically tailored to a particular scientific question.

The real advantage of low-$z$ analogues is that they allow us to study
the well-understood optical spectral region with high-S/N spectra
\citep[see][for recent
reviews]{kewleyUnderstandingGalaxyEvolution2019,maiolinoReMetallicaCosmic2019}
and it is easier to collect multi-wavelength data to interpret these
observations than at high redshift. That said, rest-UV spectroscopy of
local analogues has been hard to get \citep{2011AJ....141...37L}
although the CLASSY survey
\citep{bergCOSLegacyArchive2022,jamesCLASSYIITechnical2022a} has
recently made big strides forwards here. It has also been challenging
to acquire near-IR spectroscopy of these analogues
\citep[e.g.][]{vanziIntegralFieldNearinfrared2008,huntSpitzerViewLowMetallicity2010a,cresciIntegralfieldNearinfraredSpectroscopy2010,izotovNearinfraredSpectroscopyLarge2016a},
something that might hamper future interpretations of NIRSpec data on
$z\sim 1$--3 galaxies.

The advent of rest-frame optical spectroscopy of $z>5$ galaxies with
the James Webb Space Telescope (JWST) NIRSpec spectrograph
\citep{jakobsenNearInfraredSpectrographNIRSpec2022,ferruitNearInfraredSpectrographNIRSpec2022}
has now opened up the possibility to contrast these low-redshift
counter-parts with the real high-redshift galaxies and to better
understand how local analogues can help guide our study of high-$z$
galaxies.  But it is worth reflecting upon what analogues can be
useful for and a caution that an analogue is just that, and should not
be viewed as a replacement for studying the high redshift galaxies
directly.

To do so, it is useful to distinguish between large-scale
environmental properties, extrinsic properties (scaling with the size
of the system) and intrinsic properties. There is no real way to find
fully equivalent environmental conditions for low-z analogues as
compared to a $z\sim 8$ galaxy since the Cosmic Microwave Background
temperature is $9 (1+z)/(1+8)$ times higher and the mean density of
the Universe is $\sim 730 \left((1+z)/(1+8)\right)^3$ times denser
than today. Since the large-scale environment frequently has an
impact on the extrinsic properties of galaxies
\citep[e.g.][]{pengMassEnvironmentDrivers2010}, this might argue
against their use for comparisons. Here I will therefore focus on
intrisic quantities and in particular emission line ratios to match
samples at low and high redshift.

The public release of an Early Release Observations using NIRSpec over
the SMACS J0723.3–7327 galaxy cluster (SMACS 0723 hereafter) spurred a flurry of studies of the rest-frame optical spectra of the
five clear high-z galaxies in the data
\citep{curtiChemicalEnrichmentEarly2023,trusslerSeeingSharperDeeper2022,katzFirstInsightsISM2023,carnallFirstLookSMACS07232023,schaererFirstLookJWST2022,trumpPhysicalConditionsEmissionLine2023,rhoadsFindingPeasEarly2023,arellano-cordovaFirstLookAbundance2022,tacchellaJWSTNIRCamNIRSpec2023,taylorMetallicitiesFiveEmissionline2022},
and I will return to a comparison to some of these results further
below.  Subsequently a number of NIRSpec studies have extended
  this sample to higher redshift
  \citep[e.g.][]{bunkerJADESNIRSpecSpectroscopy2023,curtis-lakeSpectroscopicConfirmationFour2023,robertsonIdentificationPropertiesIntense2023}
  as well as more extensively covering $z\sim 1$ to $z\sim 9$
  \citep[e.g.][]{sandersDirectEbasedMetallicities2023,cameronJADESProbingInterstellar2023,shapleyJWSTNIRSpecBalmerline2023,reddyPaschenlineConstraintsDust2023,masedaJWSTNIRSpecMeasurements2023}
  and some exploration of non-star formation sources
  \citep[e.g.][]{larsonCEERSDiscoveryAccreting2023}.

The focus in many of these papers has been on the $z>6$ galaxies
  given their novelty and closeness
to the epoch of reionization, the peak of star formation in the
Universe happens between $z=1$ and $z=3$
\citep{madauCosmicStarFormationHistory2014} and this is a redshift
range where NIRSpec gives access to a rich array of near-IR
lines. These include Paschen and Brackett lines of Hydrogen, which are
less affected by extinction than optical lines and hence are powerful
references for star formation rate estimates
\citep[e.g.][]{2012ARA&A..50..531K,reddyPaschenlineConstraintsDust2023},
although this should be tempered 
by the fact that they are relatively weak: in the absence of dust
attenuation it is a useful rule of thumb that \Pa\ is about as strong
as \hd\ relative to \ha, \Pb\ similar to H$\epsilon$ and \Pg\ to
H($8\to 2$), and the Brackett lines a factor of $\sim 5$ weaker still.

However besides hydrogen lines, the near-IR has prominent \hei{},
\feii{}, and \hmol\ lines which offer complementary information on the
physical properties of galaxies to that provided by the optical lines.
\feii{} lines are frequently weak in star-forming galaxies because the
iron is locked up in dust grains, but they are found to be enhanced in
shocked regions because of the destruction of dust grains behind shock
fronts due to sputtering processes there
\citep{olivaInfraredSpectroscopySupernova1989,greenhouseNearInfraredFeIi1991}. This
has led to \feii{1.257} and in particular \feii{1.644} being used for
studies of supernovae in galaxies
\citep[e.g.][]{olivaDetectionSIVI1990,alonso-herreroUsingNearInfraredFe1997,rosenbergFeIITracerSupernova2012a,bruursemaSearchSupernovaRemnants2014}
and they are also regularly seen in AGN
\citep{mouriExcitationMechanismNearInfrared2000} and at a much weaker
level in star forming galaxies coming from photoionization
\citep{izotovNearinfraredSpectroscopyLarge2016a,cresciIntegralfieldNearinfraredSpectroscopy2010,vanziIntegralFieldNearinfrared2008,vanziIntegralFieldSpectroscopy2011}. As
we will see below, they are also seen at an interesting level in the
NIRSpec Early Release Observations (ERO) data.

The molecular hydrogen emission spectrum coming from ionized gas
offers a powerful way to characterise the physical conditions in the
warm molecular regions
\citep[e.g.][]{blackFluorescentExcitationInterstellar1987,kaplanExcitationMolecularHydrogen2017}
and will provide novel information on the nature of $z\sim 2$ galaxies
although the detection of a large set of rovibrational lines of \hmol\
will likely be challenging in most cases.

That flux calibration will be a challenge for NIRSpec has been clear
for a long time given the very small slits
\citep[e.g.][]{jakobsenNearInfraredSpectrographNIRSpec2022,ferruitNearInfraredSpectrographNIRSpec2022}.
{\new In section~\ref{sec:data} I will discuss the method I adopted to
  correct the flux calibration of the spectra which is complementary
  to those used in the literature}, I also will present the new
redshift determinations here. I will discuss the emission line
measurements in section~\ref{sec:measurements}. In
section~\ref{sec:empirical} I propose a modification to the standard
temperature sensitive abundance estimation method to reduce the
senstivity to flux calibration errors and use this to derive electron
temperatures and oxygen abundances which I compare to the literature
on these sources.  \ref{sec:photo-ionization} is dedicated to fits of
photoionization models to the fluxes of these high-z galaxies and
section~\ref{sec:source_of_ionization} has an in-depth discussion of
the ionizing sources in these galaxies which until then I will tacitly
assume to be star formation. In section~\ref{sec:comp_to_local} I
compare to local analogues and this feeds into the discussion in
section~\ref{sec:gas_phase} before I conclude in
section~\ref{sec:discussion}.

Where relevant I have adopted a Kroupa \citep{2001MNRAS.322..231K}
initial mass function and I will use a cosmology with $\Omega_m=0.3$,
$\Omega_\Lambda=0.7$ and $h=0.7$. For forbidden and helium emission
lines with rest wavelengths below 1$\mu$m I will indicate that
wavelength of the transition in \AA, while for longer wavelength lines
I will indicate the wavelength in $\mu$m. Thus \siii{9533} for the
\siii{} line at 9533.2\AA, but \hei{1.083} for the \hei{} line at
1.083$\mu$m. All wavelengths are given in vacuum. 

\section{Data}
\label{sec:data}

I will here use the JWST NIRSpec observations taken as part of the
SMACS0723 Early Release Observations (Progamme ID 2736) which have
already been discussed in detail by several authors
\citep[e.g.][]{carnallFirstLookSMACS07232023,curtiChemicalEnrichmentEarly2023,trumpPhysicalConditionsEmissionLine2023}. I
will also make some use of the deep NIRCam imaging in F090W, F150W,
F200W, F277W, F356W, and F444W released with the NIRSpec spectra, as
well as the Hubble Space Telescope (HST) images in F435W, F606W and
F814W made available by the Reionization Lensing Cluster Survey
\citep[RELICS][]{coeRELICSReionizationLensing2019}\footnote{\url{https://relics.stsci.edu/}}.
For the imaging I use the latest re-reductions provided by Gabe
  Brammer's Grizli Image Release
  v6.0\footnote{\url{https://github.com/gbrammer/grizli/blob/master/docs/grizli/image-release-v6.rst},
  but I have also used the original release of ERO images and will
  discuss the sensitivity of the results to the photometric
  calibration further below.}

Source extractor version 2.25
  \citep{bertinSExtractorSoftwareSource1996} was used to construct a
  source catalogue. For some sources Source Extractor's segmentation
  maps were not optimal for total photometry so for these six sources
  (4580, 5735, 9483, 9721, 10612, 5144) photometry was done
  manually. The effect on the results presented below is marginal
  since the colours did not change significantly. For aperture
  photometry I used a 0.4\arcsec\ diameter aperture and for total
  magnitudes I used Source Extractor's \texttt{MAG\_BEST}. To derive
  aperture corrections for the fixed apertures, I used the WebbPSF
  python package \citep{perrinSimulatingPointSpread2012} to create
  point spread functions and calculated aperture corrections assuming
  an intrinsic point source. For the manual photometry I convolved the
  images to the F444W PSF using \texttt{pypher}
  \citep{boucaudConvolutionKernelsMultiwavelength2016a} to calculate
  the appropriate convolution kernels. I chose to do this since the
  manual photometry in some cases has to use tight segmentation masks,
  but the effect of the convolution on the final photometry is $<0.1$
  magnitudes in almost all cases.

The main focus here is on the NIRSpec spectroscopy which was done with
two pointings, s007 and s008, and released as level-3 (L3) data
products on 12/07/2022. The data were obtained with two grism and
filter combinations, G235M/F170LP covered approximately
1.65\mum--3.17\mum\ and G395M/F290LP covered 2.85\mum-5.28\mum\
although close to the edges the spectra are considerably worse. I took
the spectral resolution as built from the JWST web pages, which are
those shown in \citet{jakobsenNearInfraredSpectrographNIRSpec2022}.

For the analysis here I in general combine the two pointings into one
for each grating. For the fitting of emission lines I also combine the
two gratings into a single spectrum. However for checks and tests I
also analyse each pointing or grating independently.  The data shows
various features that make analysis somewhat more challenging. In
particular the flux calibration is suspect at times and I will return
to this point below. In some cases, the spectral extraction is also
sub-optimal for redshift determination because emission lines have
less contrast. Thus throughout I made extensive use of the 2D spectra
to assess the reality of lines and also to re-extract the spectra to
have better redshift estimates.

For re-extraction, I defined a box around the region I wanted to
extract a spectrum from in the 2D spectrum and defined background
side-bands of equal width on either side of this central spectral
trace. The side-bands are subtracted off which helps remove some
structure in the spectrum as also noticed by other authors
\citep[e.g.][]{trumpPhysicalConditionsEmissionLine2023}. I also
manually edit out cosmic rays.

\subsection{Flux calibration and slit losses}
\label{sec:slitlosses}

The SMACS0723 ERO NIRSpec data have clear issues with their
spectrophotometric calibration as was noted already by the first
publication discussing the data~\citep{schaererFirstLookJWST2022} and
has been discussed repeatedly since then. Several authors have
therefore re-reduced the data
\citep{curtiChemicalEnrichmentEarly2023,trumpPhysicalConditionsEmissionLine2023}. In
particular \citet{curtiChemicalEnrichmentEarly2023} have re-reduced
the spectra using the GTO pipeline and report more physically
meaningful flux ratios. As that pipeline is not available outside the
GTO consortium, and because the work presented here was done before
the Curti et al paper appeared, I have taken a different approach,
which I will show further below gives very similar results to the
Curti et al re-analysis. 

However even with perfect reductions, it is clear that flux
calibration of NIRSpec can present significant challenges.  The
NIRSpec slit width is only 0.2 arcsec in width and as stressed
in~\citet{ferruitNearInfraredSpectrographNIRSpec2022}, this can have
significant impact on slit-losses and in particular our ability to
compare line fluxes across wide ranges in wavelength. Given this,
  I am using the originally reduced data but present a method to
  \textit{a posteriori} correct these.

To try to correct for biases in the flux calibration and mitigate and
explore the slit-loss effect, I have normalised the spectra to the
NIRCam
\citep{riekeOverviewJamesWebb2005,beichmanScienceOpportunitiesNearIR2012}
photometry. The spectra were first convolved with the NIRCam F200W,
F277W, F356W, and F444W filters\footnote{Taken from
\url{https://jwst-docs.stsci.edu/jwst-near-infrared-camera/nircam-instrumentation/nircam-filters}}.
These are then compared to the NIRCam fluxes for the sources. I
  use both the aperture corrected fluxes through a fixed 0.4\arcsec\
  aperture, and total fluxes. Unless otherwise stated, the results
  below use aperture fluxes for the normalisation. In most cases
this comparison shows a gradient with wavelength.

I then assume that the spectrum that is lost is equal to the spectrum
extracted within the slit, in other words I assume that
\begin{equation}
  \label{eq:slitloss}
  f_{\mathrm{total}}(\lambda) = f_{\mathrm{slit}}(\lambda) +
  \alpha(\lambda) f_{\mathrm{slit}}(\lambda),
\end{equation}
where ``total'' refers to the true total flux, and ``slit''
corresponds to that measured through the slit. The comparison gives us
the average of $\alpha$ over the filter, and I then assume that this
gives us an approximate measure of $\alpha$ at the pivot wavelength of
the filter. The $\alpha(\lambda_{\mathrm{pivot}})$ are then fit with a
linear function for each grism and applied as a correction to get the
final corrected spectrum.  In updating this paper a similar. but
  distinct, method was developed
  by~\citet{reddyPaschenlineConstraintsDust2023} and shown to work
  well also for their spectra.

\begin{figure}
  \centering
  \includegraphics[width=\columnwidth]{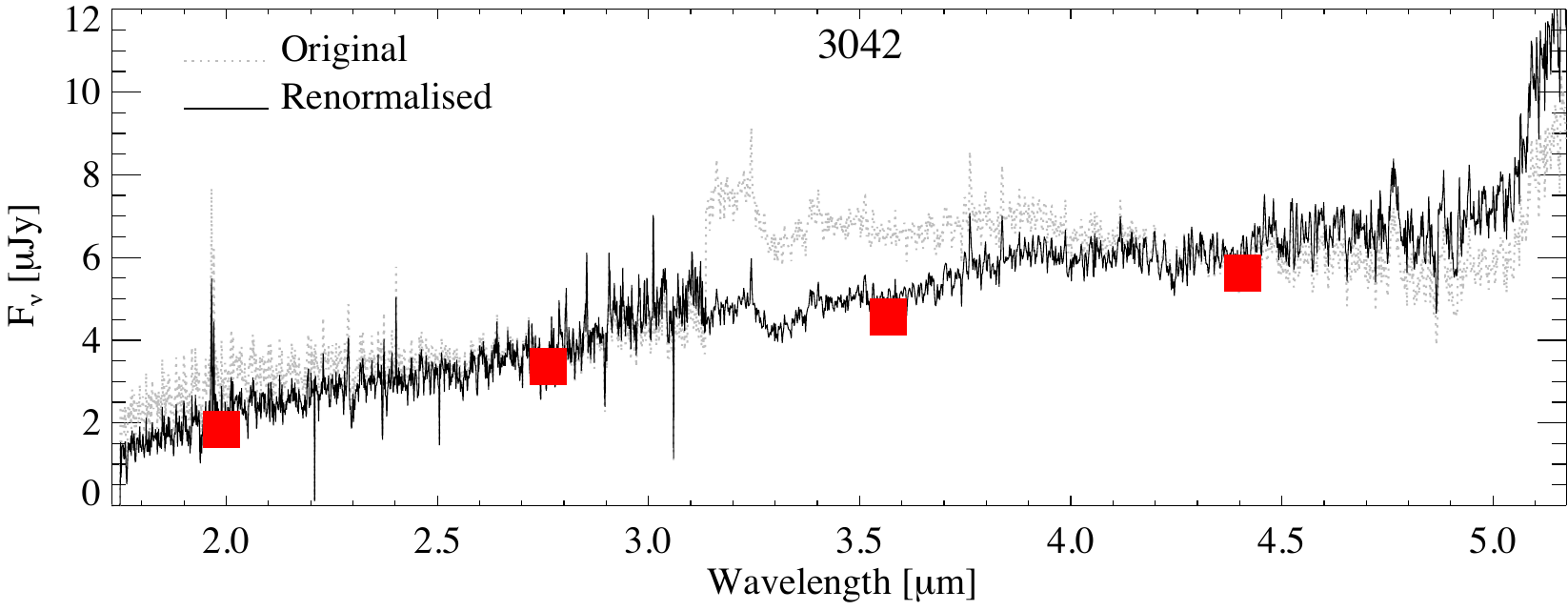}
  \caption{An illustration of the effect of flux correction of the
    NIRSpec spectrum. In this particular case, total magnitudes are
    used. The original L3 spectrum, normalised to the F277W flux is
    shown as a dotted gray line, while the re-normalised spectrum
    is shown with the solid black line. The galaxy shown here is
    3042 and this is a complex source, as discussed in the text, which
    in part is likely to be the reason for the substantial difference
    in spectral shape.}
  \label{fig:renorm_illustration}
\end{figure}
This process gives satisfactory results in most
cases. Figure~\ref{fig:renorm_illustration} shows the result of this
procedure on 3042 which, as I'll return to below, is a complex
superposition of two galaxies which is likely the cause of the
significant change in slope here. Most other galaxies show much
slighter changes, but 5144 does not yield a good solution so its
spectrum is used unmodified, while 4580 and 5735 do not have enough
fluxes in the NIRCam catalogue to be well corrected, thus these have
also been used unmodified.  Further below I will also demonstrate that
the measured Balmer ratios are physically meaningful and agree well
with the ones measured off the re-calibrated data presented
by~\citet{curtiChemicalEnrichmentEarly2023}.

Although the process superficially works well, it has, however, a
number of weaknesses: the assumption of a linear gradient with
wavelength is unjustified but can be viewed as a the lowest order
Taylor expansion of the correction function; the extrapolation to the
reddest and bluest wavelengths is significantly uncertain, and of
course the assumption of a constant spectral shape inside and outside
the slit is questionable. That said, the method
  does improve the spectral shape quite significantly in some cases and
  by combining this method with an SED fitting approach and optimised
  photometry, some of these weaknesses can be addressed
  \citep[see e.g.][]{tacchellaJWSTNIRCamNIRSpec2023}. 

It is however important to emphasise that slit-losses are unavoidable
with NIRSpec and it is therefore important to develop methods to take
this into account in the analysis. The suggestion from the JWST
documentation is to create a forward model for the light convolved
with the results for a point source. This is satisfactory if the
emission lines follow the broad-band light profile, ie.\ have constant
equivalent widths as a function of radius --- this is patently not the
case in some classes of low-z galaxies but is likely more trustworthy
at high redshift. A systematic IFU survey of galaxies with NIRSpec
will help inform this and can then be combined with forward models of the
emission line distributions
\citep[e.g.][]{cartonInferringGasphaseMetallicity2017,cartonFirstGasphaseMetallicity2018,espejosalcedoMultiresolutionAngularMomentum2022}. 

Alternatively one can limit the effect of slit-losses by constructing
analysis methods that focus on nearby emission lines. This implicitly
assumes that emission line ratios are radially constant which might be
questionable and certainly is at low redshift, however in lieu of
other information is not a bad assumption. This is the approach I will
take, and and has also been the approach taken by all the papers
  using the early ERO NIRSpec data.

\subsection{Redshift measurements}
\label{sec:redshifts}

I present here 14 high confidence redshifts, 10 of which have already
been presented in C22, one provided
in~\citet{mahlerPrecisionModelingJWST2023}, plus one lower confidence
one, coming from the same spectrum with a high confidence
redshift. The redshifts are in general obvious and have been
determined by inspection. The details of the sources are given below
in Table~\ref{tab:redshifts}. The redshift of 3042b is highly
uncertain and is discussed below.

There is however one point that warrants some discussion. This
concerns blending/multiple sources within the aperture. Like any slit
spectrograph, NIRSpec can have multiple sources fall onto the slit and
thus have multiple traces. Given the depth that NIRSpec can routinely
reach, this can introduce some challenges but as emphasised by
\citet{masedaAbundantSerendipitousEmission2019} also opportunities. In
the current dataset there are two clear cases of this: 8277 which has
two sources well-separated within the slit, and 3042 which has two
sources overlapping. For 8277 I have no secure redshift as I can only
identify one clear line for the brightest source and none for the
fainter, thus there is nothing more to be said about this here.

However, 3042 offers an interesting case study of a type that one
should watch out for in future NIRSpec studies, particularly the
deeper ones.

\begin{figure*}
  \centering
  \includegraphics[width=0.9\textwidth]{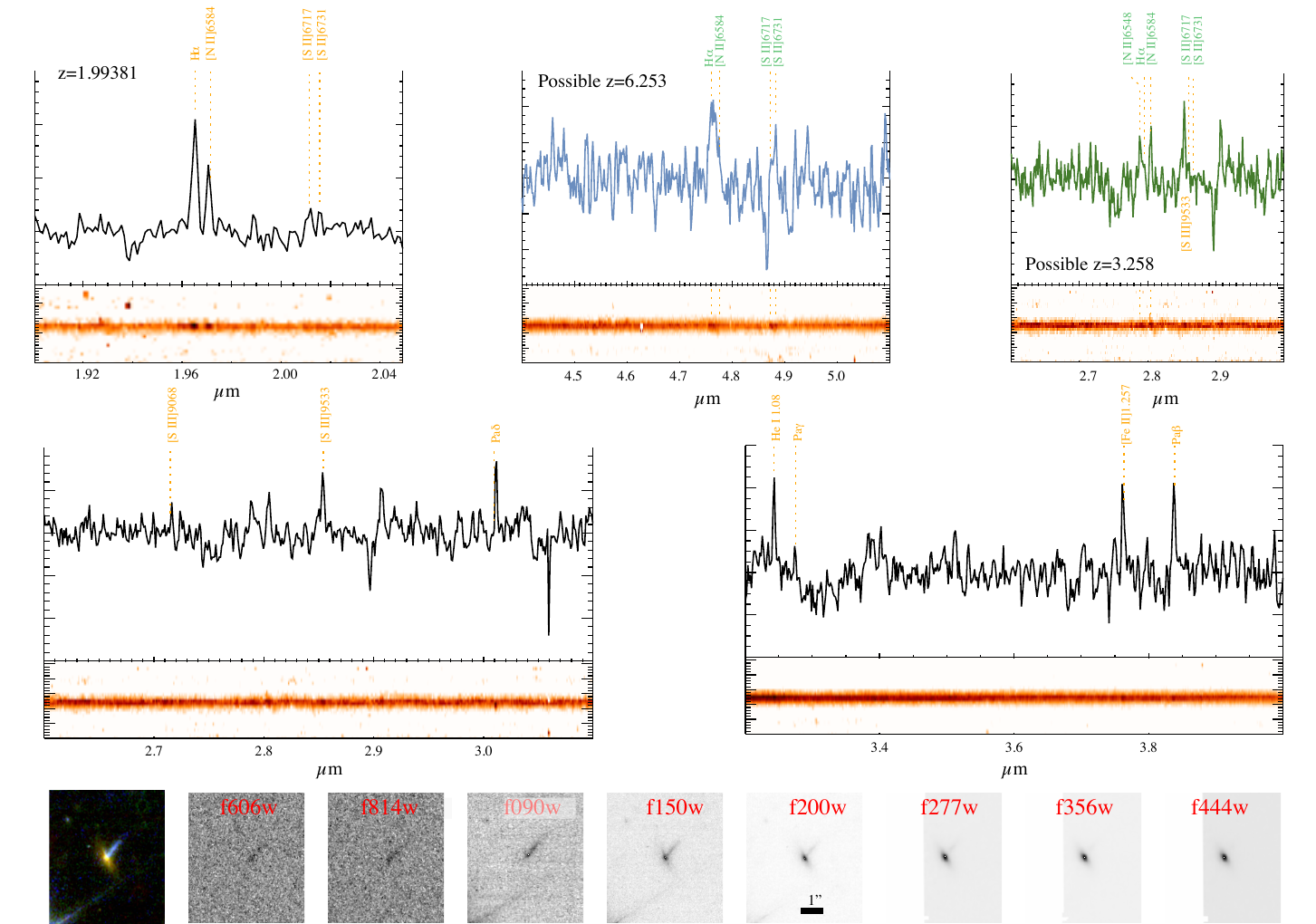}
  \caption{The overlapping galaxies in 3042. The spectrum showed here
    has the continuum subtracted. There is a very clear redshift of
    $z=1.9938$ which is demonstrated in the top left panel and
      the middle row.  This presumably corresponds to the object
    detected in the bands bluest bands (f090w and bluewards) shown in
    the bottom row. These images are all 5'' on the side. In f150w a
    second galaxy, nearly orthogonal to this stars to appear and by
    f356w only this galaxy is easily visible. The colour image on the
    left is constructed from F090W, F150W and F200W and shows this
    clearly. As discussed in the text, there are two possible
      redshifts for this source: $z=6.253$ illustrated in the top
      middle panel, and $z=3.258$ illustrated in the top right
      panel. }
  \label{fig:3042ill}
\end{figure*}

3042 is rather more complex than it might appear at first
glance. There is a clear detection of the source in ACS F606W images
and it continues to be detected through all NIRCAM images. However,
the morphology changes dramatically through F090W and into F200W as
illustrated in Figure~\ref{fig:3042ill}. The colour image on the left
on the bottom row combines the F090W, F150W, and F200W images to
demonstrate the clear colour difference between the objects. 

The simplest interpretation of this is that we have two sources
overlapping. This has been long recognised to be a challenge with deep
MUSE observations
\citep[e.g.][]{baconMUSE3DView2015,brinchmannMUSEHubbleUltra2017a,baconMUSEExtremelyDeep2021a}
where a secure redshift can be easily found, but the assignment of
this to a photometric object can be challenging, adding another axis
to the classical spectroscopic confidence assessment. This is also the
case here and this is likely to happen frequently in deep NIRSpec data
as discussed in detail
by~\citet{masedaAbundantSerendipitousEmission2019}, and clearly
multi-band JWST/HST imaging will be paramount in dechiphering these
cases.

Here the low redshift solution is clearly $z=1.9938$ with \ha,
\nii{6584}, \siii{9533}, Pa-$\delta$ and Pa-$\beta$ all clearly seen
although Pa-$\gamma$ is not very well detected as it falls at a
wavelength where the flux calibration is particularly
problematic. \feii{1.257} is also seen.  My interpretation of this is
that the elongated source detected in the ACS images is a lower
redshift galaxy and that the redshift corresponds to this.

In contrast the source seen in F150W and redwards appears to be a
higher redshift source.  There are two possible redshifts
  for this source based on emission liens. There appears to be two
  lines in the spectrum at 4.7634 and 4.8828 micron, clearly seen in
  the 2D spectrum as can be seen in the top middle panel in
  Figure~\ref{fig:3042ill}.  The separation of these two lines matches
  well \ha\ and \sii{6717,6731} at $z=6.253$ and there are not really
  any other set of lines that match equally well.  There are also no
  signs of \oiii{4959,5007}. Thus the confidence for this redshift is
  very low.  In addition there are two lines at 2.852$\mu$m and
  2.873$\mu$m also visible in the 2D spectrum. Their separation is
  well fit by \nii{6548,6584} at $z=3.258$, but again no further
  supporting lines are visible so the confidence is low.

Photometric redshifts offer a possible way to distinguish between
  these two sources. To that effect I measured the flux of the high
  redshift sources in the NIRCam images, after masking out the lower
  redshift source. These fluxes were given to the EAZY
  \citep{Brammer:TheAstrophysicalJournal:2008} photometric redshift
  code using the approach used for very faint galaxies in
  \citet{brinchmannMUSEHubbleUltra2017a}. The resulting $P(z)$ is
  broad but rules out any redshift solutions $z>4.5$ and
  $z<1.5$. While not conclusive since photo-z outliers do exist, this
  at least argues against the $z=6.253$ solution and I have there
  adopted the $z=3.258$ solution below but it is important to stress
  that this is highly uncertain.

\section{Emission line measurements}
\label{sec:measurements}

Since one of the main goals of this paper is to compare the high-z
galaxies with local counter-parts, it is preferable to use the same
analysis tools used at low-z for line measurements. Thus I have
adopted the \texttt{platefit} tool used as a basis for the MPA-JHU
database\footnote{https://wwwmpa.mpa-garching.mpg.de/SDSS/DR7/} of galaxy properties
\citep{Tremonti2004,2004MNRAS.351.1151B,brinchmannEstimatingGasMasses2013}. This
has been well tested over two decades on surveys such as VIMOS Very
Deep Survey \citep[VVDS,
e.g.][]{lamareillePhysicalPropertiesGalaxies2009} and MUSE GTO
\citep{baconMUSEExtremelyDeep2021a} and is fairly robust to
spectrophotometric calibration errors and low S/N spectra.

Briefly, to estimate the continuum, \texttt{platefit} carries out a
non-negative least squares \citep{Lawson-Hanson-73} combination of
\citet[BC03, ][]{2003MNRAS.344.1000B} simple 
stellar population models as well as a power-law attentuation of the
continuum. The best fit model is adopted and subtracted off. On the
residual spectrum a smoothed continuum is constructed by taking a
median smooth with box size 151 pixels and and additional boxcar
smooth with a window of 51 pixels. These smooth sizes were optimized
for SDSS spectra, but were found to perform satisfactorially on the
JWST spectra as well. They are too wide to handle abrupt changes in
the continuum between gratings due to calibration issues but as none
of the main lines to be measured fall close to these edges this is not
a major problem.

After also subtracting this smooth continuum, the emission lines are
fit jointly in velocity space with a common velocity width and overall
velocity offset, assuming that the line shape is Gaussian. As remarked
above, we take the line spread function from the JWST web-site as
measured pre-launch. For the MPA-JHU catalogue we tied \nii{6548} to
\nii{6584} using a theoretical ratio of $1/3$. I have updated this to
use the latest NIST value of the relative ratio of $1/2.957$ and also
tie the fluxes of \neiii{3869} and \neiii{3967} together using a
theoretical ratio of $1/3.22$. I do not tie the more widely separated
pairs of lines \oiii{4959,5007}, \siii{9069,9533}, \feii{1.257,1.644}
together as the residual flux calibration uncertainties could play
havoc with this.

The resulting fits are acceptable but the low S/N in the continuum, if
it is even detected, means that the continuum fit is poorly
constrained. This does not have a strong influence on the emission
line measurements but more care will be needed to obtain measurements
of continuum properties. In particular the method used above, and used
for the MPA-JHU catalogue, ignores the effect of a nebular continuum
although the power-law attenuation accounts for this some
extent. However, as shown by
\citet{cardosoSelfconsistentPopulationSpectral2019,pappalardoSelfconsistentPopulationSpectral2021}  this can be important
precisely for these kinds of galaxies, and a more careful analaysis
with e.g. FADO \citep{gomesFittingAnalysisUsing2017} as has been done
for extreme emission line galaxies by~\citet{bredaCharacterisationStellarContent2022a}, or
using a modified \texttt{platefit} as done by~\citet{gunawardhanaStellarPopulationsPhysical2020a} in their
analysis of the MUSE data on the Antennae, would address
this. Secondly, the BC03 models are effectively based on scaled-solar
abundance spectra and it is highly likely that the stars making up the
stellar continuum in the highest redshift galaxies have different
abundance ratios due to the different enrichment time-scales for
core-collapse and type Ia supernovae, and indeed this has been argued
to be of major importance also at redshift $z\sim2$--$3$
\citep{stromNebularEmissionLine2017,toppingMOSDEFLRISSurveyConnection2020}. For
the optical region of interest here, however, this should not be
important.

In the lower redshift galaxies, where NIRSpec covers the rest near-IR
to red optical, the \texttt{platefit} results are less satisfactory in
part due to the lower spectral resolution of the BC03 templates
used. I therefore also fit the spectrum manually using Gaussians. I
adopt joint fits of Gaussians to blended, or nearly blended, lines
such as \ciii{1907,1909}, \oii{3726,3729}, \sii{6717,6731} and a joint
triple Gaussian fit for \nii{6548,6584} and \ha. These Gaussian fits
compare well with the \texttt{platefit} results but behave better in
the near-IR part of the spectra so I'll adopt those for the analysis
of the $z<3$ galaxies.

\subsection{Noise estimates}
\label{sec:noise}

It is imperative to have good noise estimates for spectra before one
measures line fluxes and in particular try to infer physical
parameters from the line ratios which are more sensitive to noise. The
most common approach in the papers appearing thus far has been to
adopt the noise estimate from the pipeline. This appears to
underestimate the true noise in the spectrum, a fact also noted by
several other researchers
\citep[e.g.][]{rhoadsFindingPeasEarly2023,trumpPhysicalConditionsEmissionLine2023}.

We can quantify this by running \texttt{platefit} on each
individual observation separately and comparing the derived
fluxes. Concretely I selected all lines with S/N$>5$ (a cut of 7 or 10
gives comparable results), and calculated the standardized difference,
\begin{equation}
  \label{eq:norm_diff}
  \Delta = \frac{f_{007}-f_{008}}{\sqrt{\sigma_{007}^2 +
    \sigma_{008}^2}}, 
\end{equation}
where 007 and 008 refer to the individual observations and $\sigma$ is
the flux uncertainty. Under the assumption that all uncertainties are
normal, this should be distributed as a unit variance normal
distribution. Deviations from this can come from two sources. Firstly,
the uncertainty estimates themselves are uncertain and if this is
important, $\Delta$ will be distributed as a Student's t distribution
rather than a normal --- this is most easily seen in the wings of the
distribution but as only 73 flux measurements were available here,
this is impossible to assess. The other, and more relevant source is
that underestimated uncertainties tend to inflate $\Delta$. For SDSS
DR7 we found that this was indeed a concern
\citep[][BC13 hereafter]{brinchmannEstimatingGasMasses2013} and here too I find that the
uncertainties delivered by \texttt{platefit} are underestimated by a
factor of 2.75 when using the pipeline reductions. A part of this is
due to \texttt{platefit} only reporting the diagonal elements of the
covariance matrix, but when applying gaussian fits manually to the
lines I also find a discrepancy of a factor of 2 so this appears
robust. I also find a slight bias between the two observations in that
008 is slightly brighter. Given the combination and subsequent
renormalisation of the spectra, this does not matter for the results
presented here.

More directly, it is notable that the root-mean-square of the spectrum
in line-free regions is larger than expected from the pipeline noise
spectrum so I have opted to be conservative and adopt an empirical RMS
spectrum as my noise spectrum. I use a window of six pixels, calculate
the standard deviation in this sliding window. reject outliers that
are more than 3$\sigma$ deviant and recalculate the standard
deviation. The results presented below are not very sensitive to this
approach but the noise correction will become an issue again in
section~\ref{sec:discussion} below.

\subsection{Sample characteristics and source description}
\label{sec:desc}

\begin{table*}
  \centering
  \caption{The redshifts and main lines indentified in the sources
    discussed here. Most of these were already presented
    in~\citet{carnallFirstLookSMACS07232023} but 3042, 5735, 9721 and
    potentially 3042b are new here. The uncertainties on the redshifts
    are from Gaussian fits to individual lines.}
  \label{tab:redshifts}
\begin{tabular}{cccp{0.25\linewidth}p{0.25\linewidth}} \hline
Object  & Redshift & Confidence & Lines & Comments \\ \hline
  1917  & $1.2430 \pm 0.0003$ & 3.0 & \siii{9533}, \hei{1.083}, \Pa &
  \\
  3042  & $1.9934 \pm 0.0040$ & 3.0 & \ha, \sii{9068,9533}, \hei{1.083}, \Pg
 & Two overlapping galaxies with one invisible in F090W. The redshift
\citet{mahlerPrecisionModelingJWST2023}   is for the lower redshift source.  \\
  3042b & $3.258$  & 1.0 & Possible \nii{6548,6584} & Highly
                                                          uncertain redshift. \\
  4580  & 5.1727 & 2.0 & \oiii{5007}, \ha.
  & Pointed out by~\citet{mahlerPrecisionModelingJWST2023}. Only visit
    008 is used to avoid problems near \ha
  \\ 
  4590  & $8.4951 \pm 0.0021$  & 3.0 & \ciii{1909}, \oii{3727}, to
                                       \oiii{5007}
  & Clear redshift but only observation 008 used \\
  5144  & $6.3787 \pm 0.0007$ & 3.0 & \oii{3727} through to \ha &
  Only exposure 007 was used. Spectrophotometric recalibration unsatisfactory. \\
  5735  & $1.5073 \pm 0.0004$ & 3.0 & \siii{9533}, \hei{1.083}, \Pa,
                                      \Pb & Unsatisfactory recalibration \\
  6355  & $7.6640 \pm 0.0010$ & 3.0 & \neiv{2423}, \oii{3727}, to
                                    \oiii{5007} & \neiv{2423} seen in both 007 and 008.\\ 
  8140  & $5.2745 \pm 0.0013$ & 3.0 & \oii{3727} through to \ha
  & The combined spectrum has a feature at \nev{3428} but this appears
  only in 007. \\
  8506  & $2.2115 \pm 0.0004$ & 3.0 & \ha, \sii{6716,6732},
                                    \siii{9068,9533}, \Pd,
                                    \hei{1.083}, \Pg, \Pb &  \\
  9239  & $2.4624 \pm 0.0006$ & 3.0 & \ha, \sii{6716,6732},
                                    \siii{9533}, \Pd,
                                    \hei{1.083}, \Pg, \Pb,
                                    \feii{1.257} & Strong continuum so
  lines are hard to confirm in 2D image.\\
  9483  & $1.1616 \pm 0.0003$ & 3.0 & \siii{9068,9533}, \hei{1.083},
                                    \feii{1.257}, \Pg, \Pb, \Pa, \Brd,
                                        \Brg, \hei{2.058}, H$_2$ 1-0
                                    S(3), H$_2$ 1-0 S(2), H$_2$ 1-0
                                    S(1) &  the only galaxy with
                                           prominent H$_2$ emission
                                           lines. \\
  9721  & $2.1184 \pm 0.0035$ & 3.0 & \ha, \sii{6717,6731},
                                    \siii{9068,9533}, \hei{1.083},
                                    \Pg, \Pb &  \\
  9922  & $2.7412 \pm 0.0004$ & 3.0 & \hb, \oiii{4959, 5007}, \ha,
                                    \sii{6717,6731}, \siii{9068,9533},
                                        \Pd, \Pg, \Pb, \hei{1.083} &  \\
  10612 & $7.6597 \pm 0.0014$ & 3.0 & \oii{3727} through to \oiii{5007}
                                        & The highest ionization
                                          parameter galaxy \\ \hline
\end{tabular}
\end{table*}

The basic characteristics of the sources with secure redshifts are
given in Table~\ref{tab:redshifts}. Line fluxes measured for the
sources are provided in the online material. The two sources 4580 and
3042b will not be included in the emission line analysis below as they
have insufficient spectral information to be useful.

\section{Temperatures and abundances  using the empirical method}
\label{sec:empirical}

The classical way to estimate temperatures of ionized nebulae is based
on the ratio of the fluxes of auroral transitions to lower lying
transitions. This is a fairly straightforward technique and it was
already well-established when Aller wrote his influencial book
``Gaseous Nebulae'' \citep{allerGaseousNebulae1956}. It relies on the
relatively strong temperature dependence of auroral lines relative to
lower-lying levels. 

The method has been widely used in nearby galaxies and star-forming
regions although the faintness of auroral lines has always limited
their use and they are rarely detected at high redshift
\citep[see][for a
compilation]{sandersMOSDEFSurveyDirectmethod2020} and have been almost
exclusively focused on \oiii{} detections with the recent detection of
auroral \oii{7322,7332} by \citet{sandersPreviewJWSTMetallicity2023}
the exception. 

\begin{figure*}
  \centering
  \includegraphics[width=0.9\textwidth]{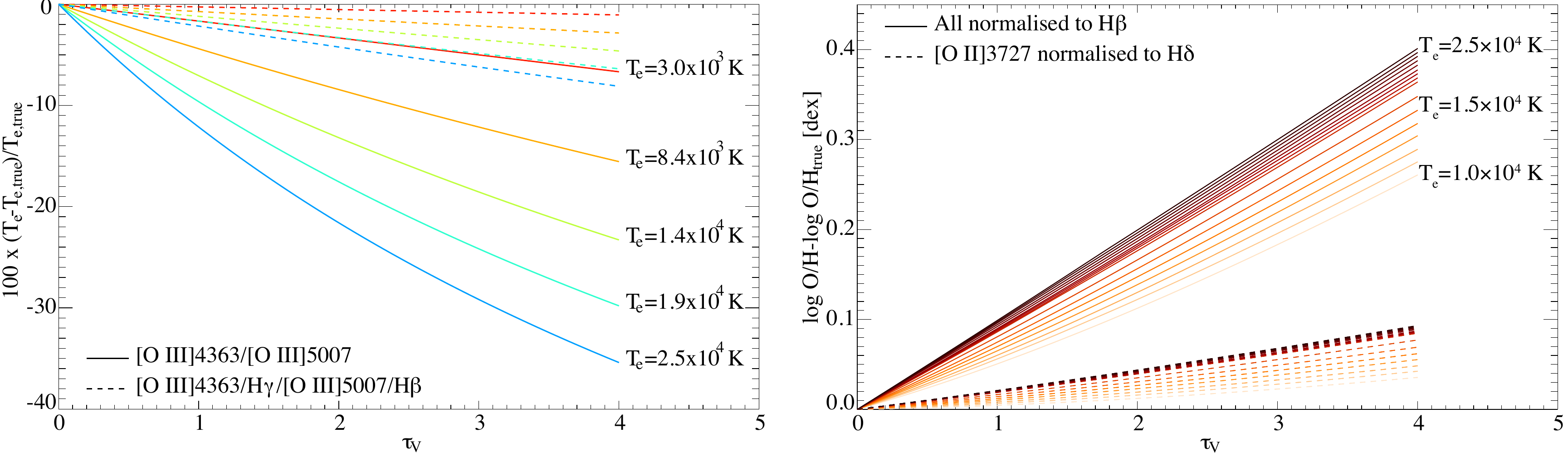}
  \caption{Left: The relative difference between estimated and true $T_e$ as
    a function of the amount of dust attenuation applied (see text for
    details). The solid lines show the results for the five true $T_e$
    values shown when applying the standard \oiii{4363}/\oiii{5007}
    ratio, while the dashed lines show the same when using the
    proposed double line ratio,
    \oiii{4363}/\hg/\oiii{5007}/\hb. Right: the difference in
    estimated $12+\log \mathrm{O/H}$ as a function of dust attenation
    for the standard method (solid lines) and the proposed robust
    method (dashed lines). }
  \label{fig:te_sensitivity}
\end{figure*}

The most widely used ratio at low redshift and the most interesting
one for the present sample is the \oiii{4363}/\oiii{5007} ratio of
doubly-ionized oxygen. This is insensitive to density over a
  range of densities likely to occur in star forming galaxies. For
  concreteness I will fix the density to $n_e=10^2\,\mathrm{cm}^{-3}$
  where relevant. The
limiting factor here is usually the detection of \oiii{4363} but also
the relative flux calibration of the spectrum, including
de-reddening. In view of the flux calibration issues with the spectra
it is therefore desirable to find a more robust estimator both for
temperatures and abundances.

We can improve the robustness by using the double, or composite, line
ratio\footnote{If the reader wants a shorthand, I offer OHOH but will
  refrain from using it.}  \oiii{4363}/\hg/\oiii{5007}/\hb\ instead of
\oiii{4363}/\oiii{5007}. Double line ratios were explored
by~\citet{evansTheoreticalModelsII1985a} from a different angle, but
they are not widely used. They do, however, offer a way to construct
diagnostic diagrams that are insensitive to dust attenuation or flux
calibration issues which can be helpful not only here but also when
combining JWST with ground-based or other space-based facilities.  I
note that this ratio was also used
by~\citet{trumpPhysicalConditionsEmissionLine2023} but they used it to
estimate \oiii{4363}/\oiii{5007} using a fixed \hb/\hg\ ratio, this
however ignores the temperature sensitivity of the \hb/\hg\ ratio and
as it is trivial to calculate the necessary emissivity ratios using
e.g.\ \texttt{pyneb} there is no need to make this approximation.

The proposed double line ratio normalises the fluxes of the oxygen
lines to their closest Balmer lines. That takes out any large-scale
flux calibration issues and in fact makes the line ratio quite
insensitive to dust attenuation as well. To illustrate this the left
panel of Figure~\ref{fig:te_sensitivity} shows the relative error we
make in estimating $T_e$ when not accounting for dust attenuation. I
have assumed here a power-law dust law with
$\tau(\lambda)\propto \lambda^{-1.3}$. I chose dust attenuation as
one example of a spectral slope error, other functional forms like a
power-law or a linear flux error give similar results and can give
errors in the opposite direction as well.

The panel shows the error made using the standard ratio of
\oiii{4363}/\oiii{5007} as solid lines, for five different true
temperatures as shown. The error here can exceed 30\%. In contrast,
the dashed lines show the same but now using the double line ratio
\oiii{4363}/\hg/\oiii{5007}/\hb. Each dashed line corresponds to the
similarly coloured solid line and it is clear that the double line
ratio is much more robust to calibration/dust uncertainties with an
error $<8$\%. This makes it particularly useful for the current
situation and indeed given the potentially serious slit-loss problems
for NIRSpec it would be advisable to use this approach in general.
For the specific example here, an iterative solution using Balmer
  lines as e.g. used by~\citet{curtiChemicalEnrichmentEarly2023}, is
  also insensitive. However, if instead of a smooth function with
  wavelength, there is an offset in flux, the iterative solution will
  not give correct results while the double line ratio method will
  continue to work.

While not of relevance here, I note in passing that a similar approach
can be used for some other $T_e$ estimators:
e.g. O\,\textsc{iii}]1661,1666 can be normalised to \heii{1640} and
\oiii{5007} to \heii{4686}, \siii{9533} to \Pd\ and \siii{6312} to
\ha, for instance.

The calculation of abundances follows from the calculation of
temperatures. The traditional way to do this is to calculate line
ratios relative to \hb\ and multiply this by the ratio of the
emissivity of the metal line to that of \hb. However, there is nothing
special about \hb\ and this works just as well when calculating ratios
to other Balmer lines. In the right panel of
Figure~\ref{fig:te_sensitivity}, I show the difference in derived
total oxygen abundance, $\mathrm{OH}=12+\log \mathrm{O/H}$ relative to
the true value as a function of the applied dust attenuation, here too
the dust attenuation is just a concrete example of a flux calibration
error. To construct this figure I fitted a linear relation to the
ratio of the ionic abundances of $\mathrm{O}^{+}$ and
$\mathrm{O}^{++}$ as a function $T_e$ for the sample of galaxies
analysed in~\citet{2008A&A...485..657B} up to $T_e=17,000$K and a
constant above that. Concretely I found
\begin{equation}
  \label{eq:no2no3}
  \frac{n\left(\mathrm{O}^{+}\right)}{n\left(\mathrm{O}^{++}\right)} =
  0.8523 - 0.3921 T_e(\oiii{},
\end{equation}
for $10\,\mathrm{kK} < T_e < 17\,\mathrm{kK}$. I used this to assign
\oii{3727} and \oiii{5007} fluxes using the emissivities of the two
transitions at the given $T_e$ and $n_e=10^2\,\mathrm{cm}^{-3}$. In
general the density will of course vary, but if the density is not
known, as in the case of these high$-z$ sources, it will introduce the
same bias/uncertainty regardless of the method. Here it suffices to
note that a change in density up or down by an order of magnitude will
change the \oii{3727}/\oiii{5007} emissivity by a factor of $<10$\%.
For the calculation shown in the right panel of
Figure~\ref{fig:te_sensitivity} I used 16 $T_e$ values, evenly
distributed between 10,000K and 25,000K.  The resulting fluxes were
then attenuated by dust and used to calculate $T_e$. For simplicity I
adopted single zone model with the temperature either derived using
the standard \oiii{} ratio or the double line ratio proposed above.

When using the double line ratio derived $T_e$, I calculated
abundances by normalising \oii{3727} to \hd, while for the standard
approach I normalised all to \hb. The panel shows clearly that the
standard method is sensitive to flux calibration errors or errors in
dust attenuation, while the proposed alternative method is more
robust. It could of course be made even more robust by normalising
\oii{3727} to H($9\to 2$) for instance, but H($9\to 2$) is rarely
detected and often strongly affected by stellar absorption so it is
not clear that it will offer an improvement in practice.

The double line ratio method involves four lines, so in principle
  the final uncertainty will increase. In reality this is however, not
  a significantly concern. Firstly, \oiii{4363} is nearly always
  weaker than the other lines and will dominate the uncertainty on the
  final oxygen abundance, and if one use Balmer lines to correct for
  dust, all the four lines are involved thus in the end any effect on
  the uncertainty is very minor.

\subsection{Application to the high-z galaxies}
\label{sec:Te_highz}

\begin{figure*}
  \centering
  \includegraphics[width=0.9\textwidth]{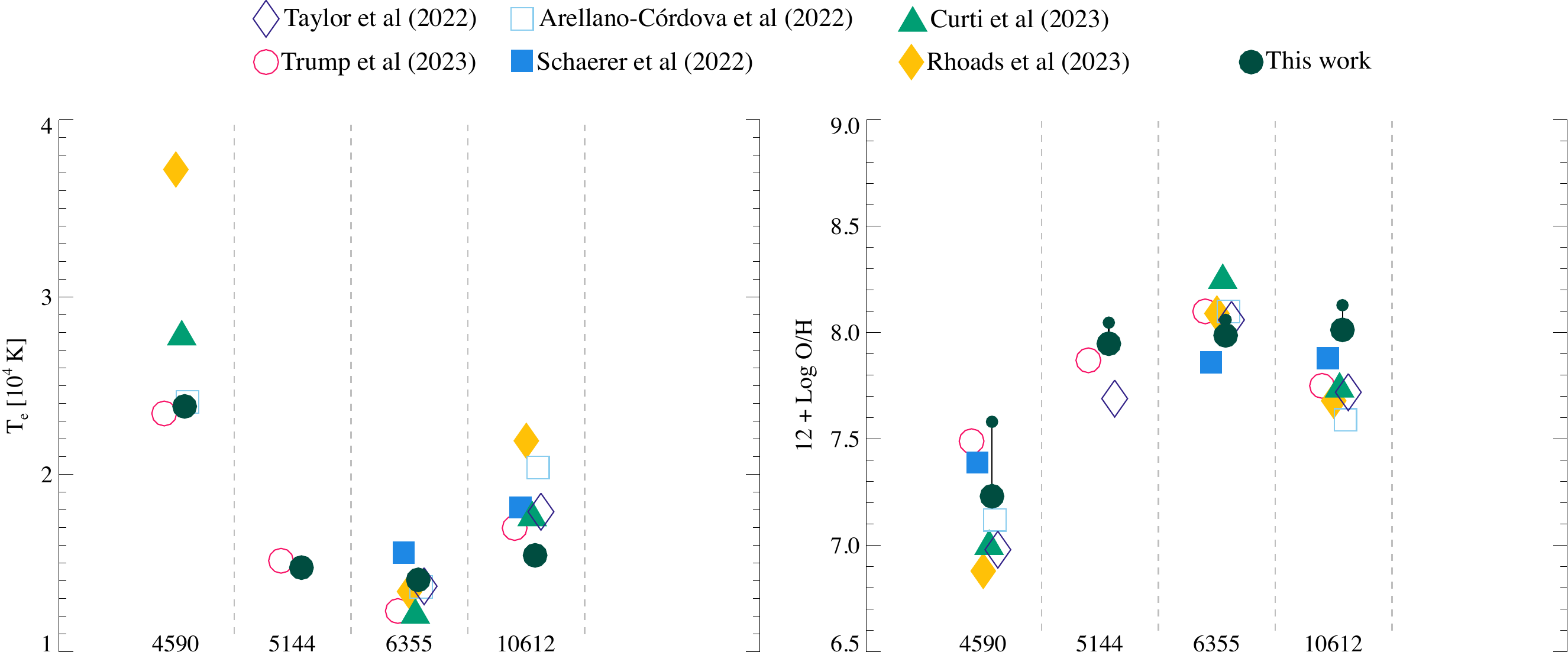}
  \caption{A comparison of literature determinations of the gas-phase
    oxygen abundance of the five high-z galaxies. }
  \label{fig:oh_comp_lit}
\end{figure*}

The oxygen abundances of the $z>5$ galaxies in the sample have been
explored extensively in the literature already.  For 4590, 6355, and
10612 we have five estimates
\citep{curtiChemicalEnrichmentEarly2023,trumpPhysicalConditionsEmissionLine2023,schaererFirstLookJWST2022,rhoadsFindingPeasEarly2023,arellano-cordovaFirstLookAbundance2022}
while \citet{trumpPhysicalConditionsEmissionLine2023} also provide
an estimate for 5144.

I apply the methods described above to estimate $T_e$(\oiii{}). To
estimate the temperature of the \oii{} emitting region we have to make
recourse to functional relations. To illustrate the importance of this
I have used three approximations: 1. set $T_e$(\oii{}) equal to
$T_e$(\oiii{}), 2. calculate $T_e$(\oii{}) using the relation proposed
by~\citet{2006A&A...448..955I}, or 3. calculate $T_e$(\oii{}) using
the relation derived
\citet{pilyuginElectronTemperaturesHighmetallicity2009}. I then 
calculate ionic abundances normalising \oii{3727} to \hd, using \hg\
leads to changes in the abundances $<0.02$dex.

The resulting $T_e$(\oiii{}) and 12+$\log $O/H values are compared to
other determinations in the literature in
Figure~\ref{fig:oh_comp_lit}. Focusing on the temperatures first, the
agreement between different authors is reasonable with some outliers
most notably for 4590 which has an apparently very strong \oiii{4363}
line. For 8140 only~\citet{trumpPhysicalConditionsEmissionLine2023}
present a temperature and abundance, I am unable to confirm their
detection of \oiii{4363} and have decided to not use the direct method
for this source. As we will see below, photoionization model fits
prefer a rather higher abundance for this source in disagreement with
Trump et al's determination.

Turning now to the oxygen abundance plot in the right panel. This
again shows decent agreement even for the extreme 4590. The effect of
the $T_e$(\oii{}) formula is shown as a solid and open smaller circle
connected by a line to the larger, filled black circle. The larger
circle corresponds to equal temperatures for the \oii{} and \oiii{}
zones, while the open corresponds to the Pilyugin et al relation and the
small filled to the Izotov et al relation. These do not have a strong
influence on the result as already noted
by~\citet{curtiChemicalEnrichmentEarly2023} although the quadratic
relation from~\citet{2006A&A...448..955I} gives discrepant results for
4590, and as \citet{rhoadsFindingPeasEarly2023} already noted this
argues for a modification of that relation. The second point is that,
like all the other publications, I have not corrected for the presence
of triply ionized oxygen as this is not expected to be a significant
contribution to the oxygen budget 
\citep{2006A&A...448..955I,arellano-cordovaFirstLookAbundance2022,bergCharacterizingExtremeEmissionline2021},
assuming of course that star formation is the dominant source of
ionization, a point I will return to in
Section~\ref{sec:source_of_ionization} below.

\section{Fitting the high-z sources with photo-ionization models}
\label{sec:photo-ionization}

In order to do a systematic study of metals in a sample of galaxies,
it is usually better to exploit strong lines rather than temperature
sensitive lines due to their better detectability. At low redshifts it
has been common to use calibrated emission line ratios as the way to
get this and this has led to numerous discussions about calibration.

However calibrated emission line ratios are a very simplistic way to
estimate emission line properties and it is much better to use all the
data and model emission line properties using fits to photoionization
models. This was first used for the SDSS
in~\citet{brinchmannPhysicalPropertiesStarforming2004a,tremontiOriginMassMetallicity2004}. It
has subsequently been used more widely
\citep{blancIZIINFERRINGGAS2015,valeasariBONDBayesianOxygen2016,fernandezNewInsightsNebular2022},
and more recently also combined with full-spectrum modeling
\citep{chevallardModellingInterpretingSpectral2016,johnsonStellarPopulationInference2021}
which also has very recently been applied to the three $z>7$ galaxies
by~\citet{tacchellaJWSTNIRCamNIRSpec2023}. It is a natural match for
the data presented here since this approach can explore a wider gamut
of model parameters than a calibrated emission line ratio method would
do \citep[see][for
discussions]{maiolinoReMetallicaCosmic2019,kewleyUnderstandingGalaxyEvolution2019}.

For the study here I used both the original code, \texttt{CL01fit},
used for the SDSS studies mentioned above and last described in B13
which makes use of the~\citet[][CL01 hereafter]{2001MNRAS.323..887C}
models, and a python re-implementation of this called
\texttt{PIModels} which has support for multiple photoionization grids
and can carry out the Bayesian analysis either using a stochastic or a
gridded approach.

In either case, the codes fit a given set of lines, $\{L_i\}$ using a
Bayesian appraoch. We calculate the log likelihood of each model,
${\cal M}(U, \xi, \tau_V, Z)$, through:
\begin{equation}
  \label{eq:Pmodel}
  \ln P({\cal M}|\{L_i\}) = -\frac{1}{2} \sum_{i \in \{L_i\}}
  \frac{ \left(f_i - A f_{\cal M}\right)^2}{\sigma_i^2} + \ln \mathrm{Pr},
\end{equation}
where $f_i$ is the flux in line $i$, $A$ is a scaling-factor and
$f_{\cal M}$ corresponds to the relevant model. $\mathrm{Pr}$ denotes
the prior on the model parameters. For most use this high-dimensional
distribution is then marginalised down to 1D or 2D probability
distribution functions (PDFs) --- for the most part I will focus on 1D
PDFs in this paper. I will present results based on both the CL01 and
\citet[][G16 hereafter]{gutkinModellingNebularEmission2016}
models. For the CL01 models I will use the \texttt{CL01fit} code,
while for G16 I use \texttt{PIFit}.

\texttt{CL01fit} is described in detail in B13. It is a grid-based
Bayesian code and evaluates equation~\eqref{eq:Pmodel} on a grid of
model parameters. This is a very fast way to do Bayesian inference and
it does not raise any issues of convergence or burn-in, but it
provides results on a fixed grid so can not adapt to very high
signal-to-noise data. \texttt{PIFit} can also adopt a gridded fit, but
the default is to use the MultiNest
\citep{ferozMULTINESTEfficientRobust2009} package to do nested
inference through the \texttt{pyMultiNest} python interface
\citep{buchnerPyMultiNestPythonInterface2016}. The code interpolates
between the models using either a multi-dimensional linear or the
Radial Basis Function interpolation from \texttt{scipy}
\citep{2020SciPy-NMeth} and also fits directly for dust attenuation if
dust attenuation is not one of the model parameters (as it is for
CL01). For the results here I use the multi-dimensional linear
interpolator as it is more robust.

The different methodologies for \texttt{CL01Fit} and \texttt{PIFit} do
not significantly impact the results. However the photoionization
model adopted does matter in this case. The CL01 model is based on
a relatively old stellar populations code \citep{1993ApJ...405..538B} as well
as an earlier version of the  Cloudy photoionization code
\citep{1998PASP..110..761F}. The model grid only goes down to $12+\log
\mathrm{O/H}=7.5$ and up to an ionization parameter of $\log U=-2$, with
incomplete sampling of the model grid beyond $\log U=-2.5$. These were
not serious limitations for application to SDSS data, but as we will
see, they are more problematic when applied to the high-z
data. However, it is still useful to apply this to demonstrate what
inferred parameters are more sensitive to the model choice. 

In contrast, the G16 models are based on up-to-date stellar population
models (an updated version of the Bruzual \& Charlot 2003
models)\nocite{2003MNRAS.344.1000B} as well as version 13.03 of Cloudy
\citep{ferland2013ReleaseCloudy2013}. They span a wider range of
metallicity down to $12 + \log \mathrm{O/H}=6.5$, as well as ionization
parameters up to $\log U=-1$. This makes them much better suited to
the present dataset and they are also used in the BEAGLE code
\citep{chevallardModellingInterpretingSpectral2016}.

\begin{figure*}
  \centering
  \includegraphics[width=0.9\textwidth]{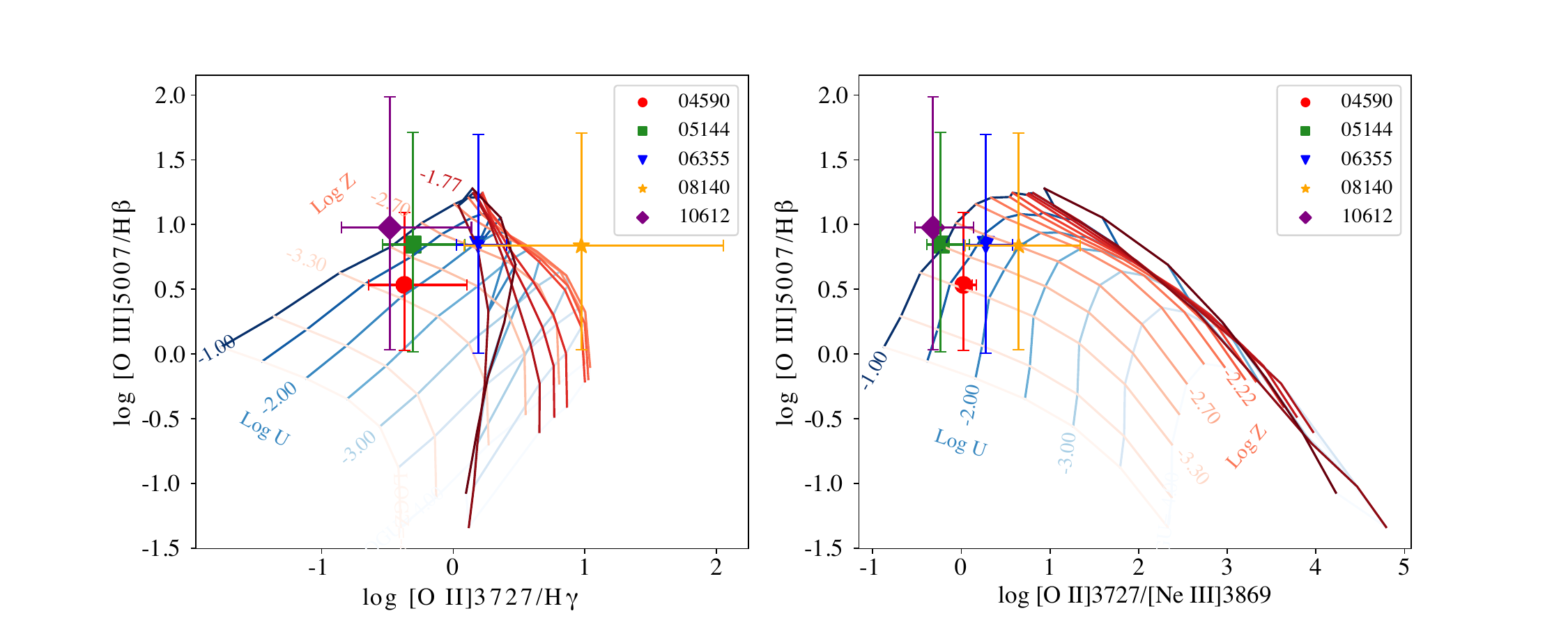}
  \caption{A comparison of the line ratios for the high-z galaxies
    compared to the G16 model grid with
    $M_{\mathrm{upper}}=100\,\mathrm{M}_\odot$ and
      $n=100\,\mathrm{cm}^{-3}$. The grid lines correspond to the
      ionization parameter, $\log U$ shown with blueish lines and
      $\log Z$ in reddish tones. }
    \label{fig:dd_against_G16}
\end{figure*}

In either case, I run the fits using \oii{3727}, \hg, \hb, and
\oiii{5007} as input fluxes. I have also run the fits with
\neiii{3869} line included and it does not change the results for G16
significantly, but for the CL01 the quality of fit is reduced so I opt
to exclude the line. I do not include the \oiii{4363} line because
photoionization models have trouble predicting this line at a
necessary accuracy
\citep[e.g.][]{dorsAnalysingDerivedMetallicities2011}. These are also
the reasons why \neiii{3869} and \oiii{4363} were not used in the
MPA-JHU metallicity fits. I also adopt flat priors on all parameters.

Before fitting, it is crucial to confirm, or not, that the model grid
covers the space probed by the galaxies to fit. This is shown in
Figure~\ref{fig:dd_against_G16} where I compare a line ratio diagram
for the G16 model against the objects. It is clear that most galaxies
fall within the model grid although 10612 is consistently on the edge
of the grid and 8140 has a somewhat discrepant \oii{3727}/\hg\ ratio
albeit with a large uncertainty. The observed line ratios were not
corrected for dust but the amount of dust allowed by the data is not
sufficient to significantly affect these ratios since they are mostly
between closely separated lines.

\begin{figure}
  \centering
  \includegraphics[width=\columnwidth]{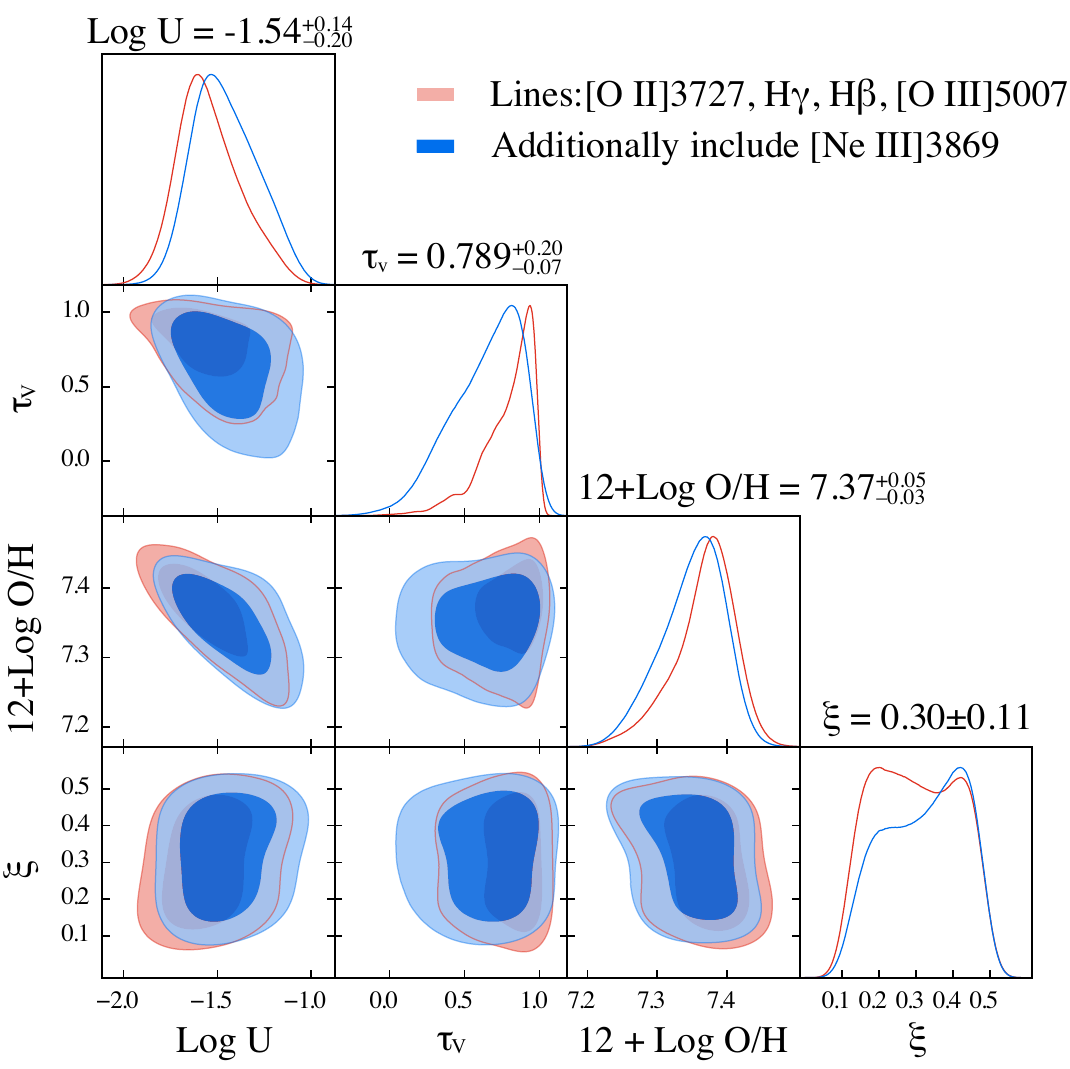}
  \caption{Triangle plots of the fits using the G16 model for
    4590. The parameters shown are the dust attenuation at $V$,
    $\tau_V$, the oxygen abundance, the dust-to-metal ratio, $\xi$,
    and the ionization parameter $\log U$. The red contours show
      the result of the fit with the default number of lines and the
      blue show the effect of adding \neiii{3869} to the fit.}
  \label{fig:triangle_G16}
\end{figure}

Figure~\ref{fig:triangle_G16} shows the result of running the G16
models on the data for 4590 and it evinces several aspects of
photoionization model fitting that are worth keeping in mind: we see a
clear correlation between dust attenuation and $\log U$. In this case
this is because the only lines that can constrain the ionization
parameter are \oii{3727} and \oiii{5007} and this means that both
increasing the ionization parameter and increasing the dust
attenuation can produce a weak \oii{3727} line, leading to the
displayed anti-correlation. This has a secondary effect on the oxygen
abudance as well. There is also a correlation between
$12+\log \mathrm{O/H}$ and the dust-to-metal ratio, $\xi$. This effect
is discussed in B13 and is due to a balance between reduced gas-phase
oxygen when $\xi$ is increased and an increased heating, see B13 for
details. With the provided lines, $\xi$ is pretty much unconstrained,
but it is an important free parameter for photoionization fits as its
value is basically unconstrained at high redshift. The plot also
  illustrates the effect of including (blue) or not (red) the
  \neiii{3869} line in the fit. In this case, the effect is very
  marginal and that is also the case for 6355, 8140 and 10612, however
  for 5144 the inclusion of \neiii{3869} helps rule out
  high-metallicity solutions.

\begin{figure*}
  \centering
  \includegraphics[width=\textwidth]{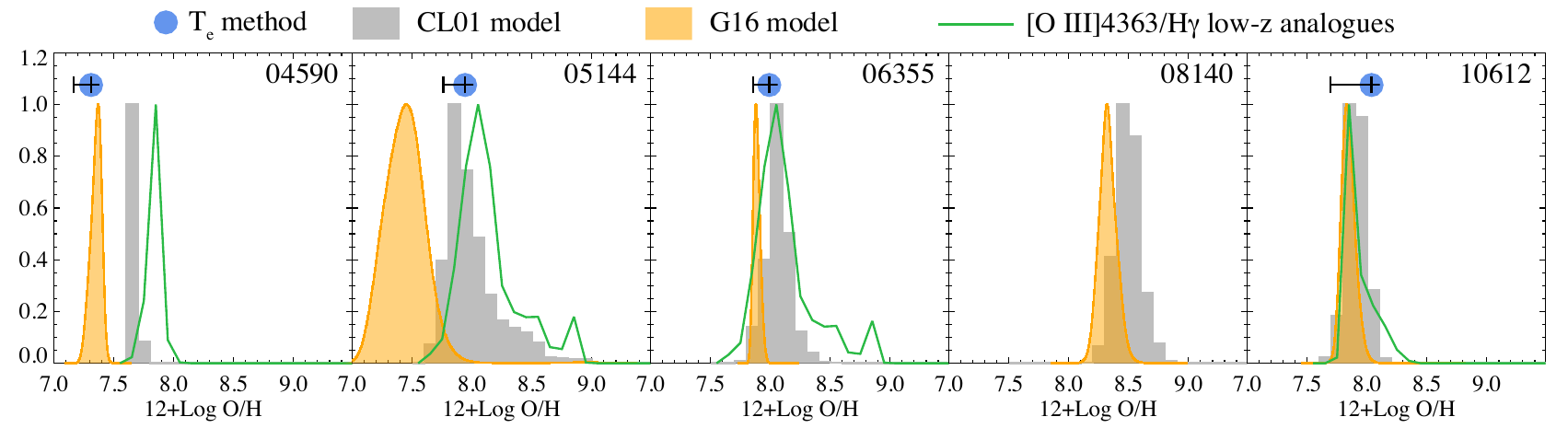}
  \caption{The resulting constraints on $12 +\log \mathrm{O/H}$ from
    fitting CL01 and G16 models to the high-redshift galaxies. There
    is one panel per galaxy and the PDF from the CL01 fit is shown as
    a filled gray histogram, while the G16 fit are shown as a filled
    orange histogram. The direct abundance estimate discussed in
    section~\ref{sec:Te_highz} is overplotted as filled disks with
    error bars. The oxygen abundance distribution of the
    $\oiii{4363}/\hg$ analogues discussed in
    section~\ref{sec:comp_to_local} are shown as a solid green line
    and discussed further in section~\ref{sec:comp_to_local}.}
  \label{fig:cl01_G16_OH_results}
\end{figure*}

Figure~\ref{fig:cl01_G16_OH_results} shows the resulting constraints
on $12+\log \mathrm{O/H}$ from fitting the CL01 (gray filled
histogram) and G16 (orange filled histogram) models to the high-$z$
galaxies. In the case of 4590 the CL01 PDF goes right up against the low
metallicity edge and the fit is overall rather poor.  In contrast the
other four galaxies are best fit with oxygen abundances well inside
the model grid. The G16 model fits overall better but for 6355, 8140,
and 10612 the resulting $12+\log \mathrm{O/H}$ constraints are in very
good agreement with those found using CL01. 

The direct oxygen abundances calculated in section~\ref{sec:Te_highz}
are shown as the blue symbols with errorbars. We see that this agrees
well with the G16 models. This is consistent with the findings
of~\citet{dorsAnalysingDerivedMetallicities2011} who also found that
when photoionization models are considered in full, they provide
results that are in good agreement with the direct method, at least at
low metallicities.

Finally, the green lines show the distribution of oxygen abundances
(dervied using the CL01 model) of the low-$z$ counterparts discussed
in section~\ref{sec:comp_to_local}. I will return to a discussion of
this further below, but first it is pertinent to take a critical look
at the source of ionization in the galaxies.

\section{The source of ionization in the galaxies}
\label{sec:source_of_ionization}

Until now I have tacitly assumed that all galaxies in the sample have
emission lines whose ionization source is dominated by
star-formation. That is however, a rather strong assumption so it is
important to underpin this as much as possible.

\subsection{The $z>5$ galaxies}
\label{sec:zgt5_ionization}

The source of ionization in the high-z galaxies has been extensively
discussed in the literature already. It is, however, problematic to
distinguish between star-bursts and AGN at low metallicities
(\citet{2006MNRAS.371.1559G}, see also \citet{nakajimaDiagnosticsPopIIIGalaxies2022}),
thus it is perhaps not surprising that all conclude that the data are
consistent with star formation.

Here I will instead focus on the UV lines detected in 4590 and 6355 to
complement the literature studies using optical lines. Starting first
with 4590, it shows \ciii{1907,1909} in emission as also noted
by~\citet{arellano-cordovaFirstLookAbundance2022}. This line is
ubiquitous in AGN, but it is also commonly seen in star-forming
galaxies
\citep{starkUltravioletEmissionLines2014,rigbyIIIEmissionStarforming2015,masedaMUSEHubbleUltra2017a,ravindranathSemiforbiddenIIIL19092020},
in particular in those that show high intensity star bursts, so it is
perhaps not surprising that it is seen here but as we will see, it
does offer some extra information on the source.

\begin{figure*}
  \centering
  \includegraphics[width=0.8\textwidth]{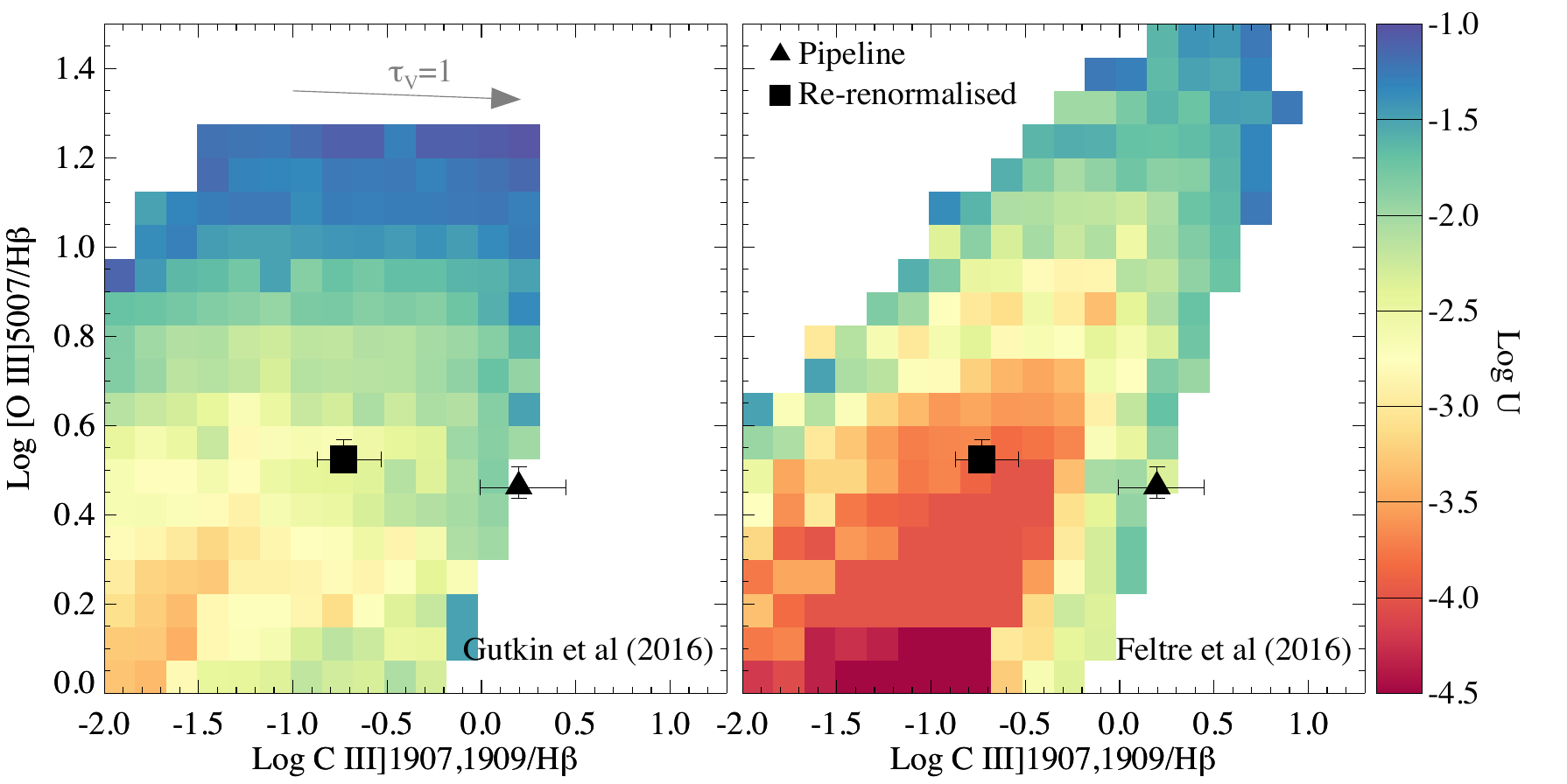}
  \caption{A comparison of model predictions for the line ratio
    \ciii{1907,1909}/\hb\ versus \oiii{5007}/\hb\ against the observed
    ratio in 4590. Left: the colour scale shows models for
    star-forming galaxies by Gutkin et al (2016) coloured by the
    average $\log U$ in each bin. The effect of applying a dereddening
    of $A_V=1$ magnitude to the observed line ratios is indicated by
    the arrow. Right: the same, but for the models
    for AGN by Feltre et al (2016). }
  \label{fig:g16_f16_comp}
\end{figure*}
Unfortunately \ciii{1907,1909} is at the very blue of the spectral range
and the renormalisation of flux that has been used here is not
reliable at the edges. Thus I have also re-measured the flux of
\ciii{1907,1909} on the uncorrected spectrum. There  are unfortunately
no nearby lines to normalise to, thus I normalise to \hb\ which is the
stronger recombination line in the spectrum.

From the unmodified spectrum I find a \ciii{1907,1909}/\hb ratio of
$1.575_{-0.59}^{+1.21}$, and from the corrected spectrum I find
  $0.186^{+0.11}_{-0.05}$. Despite the substantial uncertainties, we
can robustly conclude that the \ciii{1907,1909} flux is comparable to,
but somewhat lower than, that of \hb. This is indicative of
either a non-thermal ionization source or a moderate to high
ionization parameter in a star-burst. I illustrate this in
Figure~\ref{fig:g16_f16_comp}. This compares the measured
\ciii{1907,1909}/\hb\ ratios to
the~\citet{gutkinModellingNebularEmission2016} models for star-forming
galaxies on the left, and to the \citet{feltreNuclearActivityStar2015}
models for AGNs on the right. The 2D histograms show the average
ionization parameter, $\log U$, in each bin. Overplotted on these
diagrams are the two estimates of the \ciii{1907,1909}/\hb\
ratio. They are clearly discrepant, implying significant systematic
uncertainties, but both measurements are consistent with the models
although no distinction can be made between the two sources of
ionization. A reddening vector corresponding to
  $\tau_V=1$ is included in the left-hand panel. Clearly only a modest
  amount of dust can be accommodated before the data fall outside the
  model grid, in good agreement with
e.g. \citet{curtiChemicalEnrichmentEarly2023}.

The equivalent width of \ciii{1907,1909} is potentially a better
discriminator of galaxy properties than these line ratios
\citep[e.g.][]{jaskotPhotoionizationModelsSemiforbidden2016}, however
the flux calibration problem and non-detection of the continuum means
that this is very challenging and the data have no real constraining
power. Formally, using the total F150W flux to estimate the continuum,
the equivalent width is between 1 and 30\AA, depending on the
normalisation used and fitting method adopted. That is the entire
range spanned by star forming galaxies
\citep[e.g.][]{rigbyIIIEmissionStarforming2015} and does not require
an AGN.

\begin{figure}
  \centering
  \includegraphics[width=\columnwidth]{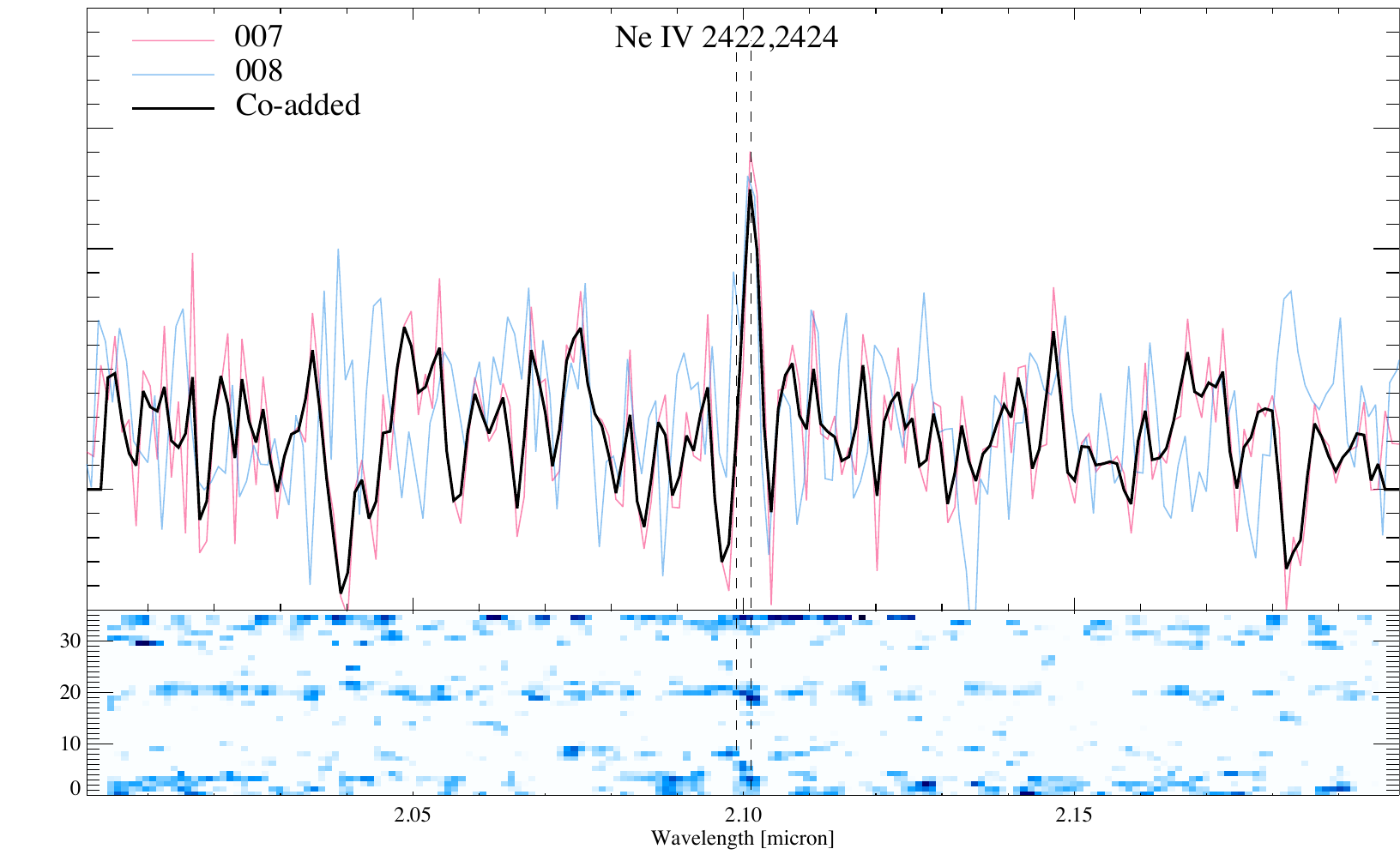}
  \caption{The \neiv{2422,2424} line in 6355. The top panel shows the
    coadded spectrum (solid black) and the individual exposures. It is
    clear that \neiv{2422,2424} is detected consistently in both
    observations. It is also clearly seen in the 2D spectrum below.}
  \label{fig:neiv}
\end{figure}
Turning now to 6355, this is an interesting case because it shows a
clear detection of \neiv{2422,2424}, as shown in
Figure~\ref{fig:neiv}. The creation of Ne$^{+++}$ requires photons
with energies $>63.45$eV and it is therefore seen only in very
energetic environments. It is commonly used as a density indicator in
bright planetary nebulae
\citep[e.g.][]{keenanEmissionLinesNeIV1998,allerRemarkablyHighExcitation1999}
and it is also seen regularly in AGNs
\citep[e.g.][]{teraoMultilineAssessmentNarrowline2022}. It is not a
line associated with star forming regions, although other high
ionization lines are seen in star-forming galaxies, albeit at low
levels at both low and higher redshifts
\citep[e.g.][]{2000ApJ...531..776G,2012MNRAS.421.1043S,starkUltravioletEmissionLines2014,bergWindowEarliestStar2018,nanayakkaraExploringHeII2019,izotovLargeBinocularTelescope2021,bergCharacterizingExtremeEmissionline2021}.

The flux in the line is rather substantial although like with the
\ciii{1907,1909} line in 4590, there is substantial uncertainty in the
flux calibration at the blue edge. Taking the measurements at face
value, the flux is $1.45\times 10^{-19}$erg/s/cm$^2$, which gives a
\neiv{2422,2424}/\hb\ $= 0.55$. I am not aware of a suitable
comparison sample, but I note
that~\citet{teraoMultilineAssessmentNarrowline2022} find that
\ciii{1907,1909} is $\sim 2$ times stronger than \neiv{2422,2424} in
their sample of radio galaxies, and at low metallicity
\ciii{1907,1909} can easily be comparable to \hb, so this ratio seems
entirely reasonable. 

In conclusion it seems that 6355 is hosting a narrow-line AGN in a
galaxy with stellar mass $<10^9\,\mathrm{M}_\odot$ at $z=7.665$. To my
knowledge this is the lowest mass galaxy to host a narrow-line AGN at
these redshifts, and clearly a source that warrants a closer
assessment. In particular the \neiv{} line is weak so its significance
must be considered somewhat tentative. The optical line ratios do
  not show clear signs of AGN activity, but since the sensitivity of
  \neiv{2422,2424} to AGN activity is much stronger than the optical
  lines this should not be taken as a counter-argument. In a galaxy
  where significant star formation is taking place, the main optical
  line ratios might be insensitive to a low level of AGN activity
  which may, however, be detectable in high ionization lines
  \citep[c.f.][]{2012MNRAS.421.1043S}. 

Neither 5144 nor 10612 show any lines directly indicative of AGN
activity but I will revisit this when comparing to local analogues
below. For line ratio diagrams I point the reader to the fine
presentations already in the literature
\citep{trusslerSeeingSharperDeeper2022,katzFirstInsightsISM2023,trumpPhysicalConditionsEmissionLine2023,rhoadsFindingPeasEarly2023,curtiChemicalEnrichmentEarly2023},
all of whom do a great job of comparing the data to various comparison
samples.

Turning now to 8140, this actually shows a strong \nev{3426} line in
the co-added spectrum. This is however not visible in 008 and there is
no \heii{4686}, as well as a very low \nii{6584}/\ha\ ratio so this
should also be considered a star-formation dominated source with the
\nev{3426} detection considered spurious.

In the high-z sample we therefore appear to have 1 narrow-line AGN out
of 5, which is an intriguingly high fraction but with the sample as
small as it is, no firm conclusions should be drawn. It is however
important to keep an eye out for these in future NIRSpec campaigns.

\subsection{The $z<5$ galaxies}
\label{sec:zlt5_ionization}

Turning next to the lower redshift galaxies, the question of
ionization source is even harder to answer. There are lines that
are clearly diagnostic of AGN activity in the near-IR such as the
\heii{1.083} line or high-ionization S or Si lines \citep[see
e.g.][]{riffel82MmSpectral2006}, however these are fairly weak lines and
none are seen in the spectra of the current sample galaxies. However
the near-IR also has \feii{} lines that have long been used as shock
tracers
\citep{olivaInfraredSpectroscopySupernova1989,olivaDetectionSIVI1990},
as well as a rich spectrum of \hmol\ emission lines which can be used to
gain insight in the warm neutral gas and photon-dominated regions in
galaxies
\citep[e.g.][]{blackFluorescentExcitationInterstellar1987,kaplanExcitationMolecularHydrogen2017}.

\begin{table*}
  \centering
  \caption{Line luminosities, not corrected for lensing, for the $z<5$
    galaxies in the sample. For each galaxy three numbers are given:
    the luminosity in solar luminosities from the normalisation to the
    fixed aperture flux, the luminosity from the spectrum normalised
    to the total magnitude and the signal-to-noise of the line. }
\begin{tabular}{l|rrr|rrr|rrr|rrr}
\hline \multicolumn{1}{c|}{Line} & \multicolumn{3}{c}{ 1917 }  & \multicolumn{3}{c}{ 3042 }  & \multicolumn{3}{c}{ 5735 }  & \multicolumn{3}{c}{ 8506 }  \\ 
 &  \multicolumn{1}{c}{$L_\mathrm{aper}$}  & \multicolumn{1}{c}{$L_\mathrm{total}$}  & S/N  &  \multicolumn{1}{c}{$L_\mathrm{aper}$}  & \multicolumn{1}{c}{$L_\mathrm{total}$}  & S/N  &  \multicolumn{1}{c}{$L_\mathrm{aper}$}  & \multicolumn{1}{c}{$L_\mathrm{total}$}  & S/N  &  \multicolumn{1}{c}{$L_\mathrm{aper}$}  & \multicolumn{1}{c}{$L_\mathrm{total}$}  & S/N \\ 
 &  \multicolumn{1}{c}{L$_\odot$} &  \multicolumn{1}{c}{L$_\odot$}  &   &  \multicolumn{1}{c}{L$_\odot$} &  \multicolumn{1}{c}{L$_\odot$}  &   &  \multicolumn{1}{c}{L$_\odot$} &  \multicolumn{1}{c}{L$_\odot$}  &   &  \multicolumn{1}{c}{L$_\odot$} &  \multicolumn{1}{c}{L$_\odot$}  &  \\ \hline 
\ha  &  $\cdots$ &  $\cdots$ & $\cdots$    &  12.5 &   58.8 &  27.11    &  $\cdots$ &  $\cdots$ & $\cdots$    &  227.2 &   2104.4 &  291.65  \\ 
\nii{6584}  &  $\cdots$ &  $\cdots$ & $\cdots$    &  5.9 &   27.4 &  14.73    &  $\cdots$ &  $\cdots$ & $\cdots$    &  15.0 &   139.1 &  24.69  \\ 
\sii{6717}  &  $\cdots$ &  $\cdots$ & $\cdots$    &  1.9 &   9.6 &  4.76    &  $\cdots$ &  $\cdots$ & $\cdots$    &  8.8 &   77.9 &  10.73  \\ 
\sii{6731}  &  $\cdots$ &  $\cdots$ & $\cdots$    &  2.2 &   11.3 &  5.72    &  $\cdots$ &  $\cdots$ & $\cdots$    &  6.0 &   52.4 &  7.95  \\ 
\siii{9068}  &  $\cdots$ &  $\cdots$ & $\cdots$    &  $\cdots$ &  $\cdots$ & $\cdots$    &  7.0 &   19.1 &  5.44    &  16.8 &   138.9 &  16.52  \\ 
\siii{9533}  &  1.6 &   4.9 &  13.09    &  3.8 &   14.3 &  7.61    &  29.6 &   78.1 &  17.55    &  37.3 &   299.4 &  45.63  \\ 
\Pd  &  $\cdots$ &  $\cdots$ & $\cdots$    &  $\cdots$ &  $\cdots$ & $\cdots$    &  $\cdots$ &  $\cdots$ & $\cdots$    &  5.5 &   43.7 &  2.63  \\ 
\hei{1.083}  &  1.0 &   2.8 &  5.78    &  3.0 &   9.7 &  4.51    &  21.8 &   57.5 &  10.65    &  28.6 &   225.1 &  25.51  \\ 
\Pg  &  0.3 &   0.9 &  3.46    &  1.1 &   4.2 &  1.70    &  $\cdots$ &  $\cdots$ & $\cdots$    &  10.3 &   81.0 &  6.14  \\ 
\feii{1.257}  &  $\cdots$ &  $\cdots$ & $\cdots$    &  2.6 &   8.5 &  3.90    &  4.0 &   8.5 &  1.71    &  2.7 &   23.0 &  1.88  \\ 
\Pb  &  $\cdots$ &  $\cdots$ & $\cdots$    &  3.0 &   11.4 &  2.94    &  12.2 &   31.1 &  5.27    &  14.7 &   115.9 &  13.15  \\ 
\Pa  &  1.4 &   3.9 &  5.13    &  $\cdots$ &  $\cdots$ & $\cdots$    &  12.3 &   31.9 &  5.23    &  $\cdots$ &  $\cdots$ & $\cdots$  \\ 
\Brd  &  0.3 &   0.7 &  3.10    &  $\cdots$ &  $\cdots$ & $\cdots$    &  5.2 &   13.5 &  1.74    &  $\cdots$ &  $\cdots$ & $\cdots$  \\ 
\hmol\ 1-0 S(3)  &  $\cdots$ &  $\cdots$ & $\cdots$    &  $\cdots$ &  $\cdots$ & $\cdots$    &  $\cdots$ &  $\cdots$ & $\cdots$    &  $\cdots$ &  $\cdots$ & $\cdots$  \\ 
\hmol\ 1-0 S(2)  &  $\cdots$ &  $\cdots$ & $\cdots$    &  $\cdots$ &  $\cdots$ & $\cdots$    &  $\cdots$ &  $\cdots$ & $\cdots$    &  $\cdots$ &  $\cdots$ & $\cdots$  \\ 
\hei{2.058}  &  $\cdots$ &  $\cdots$ & $\cdots$    &  $\cdots$ &  $\cdots$ & $\cdots$    &  $\cdots$ &  $\cdots$ & $\cdots$    &  $\cdots$ &  $\cdots$ & $\cdots$  \\ 
\hmol\ 1-0 S(1)  &  $\cdots$ &  $\cdots$ & $\cdots$    &  $\cdots$ &  $\cdots$ & $\cdots$    &  $\cdots$ &  $\cdots$ & $\cdots$    &  $\cdots$ &  $\cdots$ & $\cdots$  \\ 
\Brg  &  $\cdots$ &  $\cdots$ & $\cdots$    &  $\cdots$ &  $\cdots$ & $\cdots$    &  $\cdots$ &  $\cdots$ & $\cdots$    &  $\cdots$ &  $\cdots$ & $\cdots$  \\ 
\hline
& \multicolumn{3}{c}{ 9239 }  & \multicolumn{3}{c}{ 9483 }  & \multicolumn{3}{c}{ 9721 }  & \multicolumn{3}{c}{ 9922 }  \\ 
 &  \multicolumn{1}{c}{$L_\mathrm{aper}$}  & \multicolumn{1}{c}{$L_\mathrm{total}$}  & S/N  &  \multicolumn{1}{c}{$L_\mathrm{aper}$}  & \multicolumn{1}{c}{$L_\mathrm{total}$}  & S/N  &  \multicolumn{1}{c}{$L_\mathrm{aper}$}  & \multicolumn{1}{c}{$L_\mathrm{total}$}  & S/N  &  \multicolumn{1}{c}{$L_\mathrm{aper}$}  & \multicolumn{1}{c}{$L_\mathrm{total}$}  & S/N \\ 
 &  \multicolumn{1}{c}{L$_\odot$} &  \multicolumn{1}{c}{L$_\odot$}  &   &  \multicolumn{1}{c}{L$_\odot$} &  \multicolumn{1}{c}{L$_\odot$}  &   &  \multicolumn{1}{c}{L$_\odot$} &  \multicolumn{1}{c}{L$_\odot$}  &   &  \multicolumn{1}{c}{L$_\odot$} &  \multicolumn{1}{c}{L$_\odot$}  &  \\ \hline 
\ha  &  30.1 &   772.4 &  88.47    &  $\cdots$ &  $\cdots$ & $\cdots$    &  907.7 &   920.4 &  210.67    &  127.6 &   1249.2 &  384.05  \\ 
\nii{6584}  &  13.0 &   332.1 &  44.66    &  $\cdots$ &  $\cdots$ & $\cdots$    &  35.4 &   35.9 &  10.53    &  4.2 &   41.0 &  16.16  \\ 
\sii{6717}  &  2.8 &   65.4 &  7.26    &  $\cdots$ &  $\cdots$ & $\cdots$    &  83.1 &   67.5 &  14.31    &  4.1 &   40.3 &  11.82  \\ 
\sii{6731}  &  3.7 &   82.5 &  9.07    &  $\cdots$ &  $\cdots$ & $\cdots$    &  61.0 &   46.6 &  11.16    &  3.9 &   37.6 &  11.22  \\ 
\siii{9068}  &  $\cdots$ &  $\cdots$ & $\cdots$    &  96.8 &   122.6 &  23.90    &  37.7 &   38.2 &  10.96    &  7.4 &   71.8 &  11.27  \\ 
\siii{9533}  &  7.1 &   118.8 &  10.18    &  215.5 &   273.0 &  50.58    &  56.4 &   55.5 &  12.60    &  16.1 &   153.9 &  22.98  \\ 
\Pd  &  1.5 &   25.5 &  2.34    &  $\cdots$ &  $\cdots$ & $\cdots$    &  $\cdots$ &  $\cdots$ & $\cdots$    &  2.6 &   25.3 &  3.86  \\ 
\hei{1.083}  &  4.0 &   64.1 &  5.99    &  107.4 &   132.2 &  24.31    &  45.4 &   46.0 &  5.17    &  10.8 &   106.8 &  15.06  \\ 
\Pg  &  2.2 &   39.6 &  4.98    &  49.8 &   64.9 &  12.20    &  31.2 &   31.6 &  1.49    &  3.8 &   34.2 &  5.41  \\ 
\feii{1.257}  &  2.2 &   37.1 &  3.49    &  68.3 &   83.0 &  14.34    &  $\cdots$ &  $\cdots$ & $\cdots$    &  $\cdots$ &  $\cdots$ & $\cdots$  \\ 
\Pb  &  5.6 &   86.0 &  9.48    &  134.5 &   171.0 &  28.81    &  20.9 &   21.2 &  3.33    &  4.2 &   42.1 &  5.14  \\ 
\Pa  &  $\cdots$ &  $\cdots$ & $\cdots$    &  377.1 &   478.0 &  43.17    &  $\cdots$ &  $\cdots$ & $\cdots$    &  $\cdots$ &  $\cdots$ & $\cdots$  \\ 
\Brd  &  $\cdots$ &  $\cdots$ & $\cdots$    &  $\cdots$ &  $\cdots$ & $\cdots$    &  $\cdots$ &  $\cdots$ & $\cdots$    &  $\cdots$ &  $\cdots$ & $\cdots$  \\ 
\hmol\ 1-0 S(3)  &  $\cdots$ &  $\cdots$ & $\cdots$    &  28.8 &   35.7 &  3.34    &  $\cdots$ &  $\cdots$ & $\cdots$    &  $\cdots$ &  $\cdots$ & $\cdots$  \\ 
\hmol\ 1-0 S(2)  &  $\cdots$ &  $\cdots$ & $\cdots$    &  16.7 &   18.6 &  1.69    &  $\cdots$ &  $\cdots$ & $\cdots$    &  $\cdots$ &  $\cdots$ & $\cdots$  \\ 
\hei{2.058}  &  $\cdots$ &  $\cdots$ & $\cdots$    &  22.2 &   28.1 &  1.74    &  $\cdots$ &  $\cdots$ & $\cdots$    &  $\cdots$ &  $\cdots$ & $\cdots$  \\ 
\hmol\ 1-0 S(1)  &  $\cdots$ &  $\cdots$ & $\cdots$    &  22.1 &   26.5 &  2.82    &  $\cdots$ &  $\cdots$ & $\cdots$    &  $\cdots$ &  $\cdots$ & $\cdots$  \\ 
\Brg  &  $\cdots$ &  $\cdots$ & $\cdots$    &  28.0 &   39.5 &  3.04    &  $\cdots$ &  $\cdots$ & $\cdots$    &  $\cdots$ &  $\cdots$ & $\cdots$  \\ 
\end{tabular}
  \label{tab:zlowzlum}
\end{table*}

Starting with the more classical optical line ratios, 3042 has a
\nii{6584}/\ha\ line ratio $0.64\pm0.03$ above that normally seen in
star forming regions, but in contrast the dereddened \siii{9533}/\ha\
ratio ($0.11\pm 0.02$) matches very well the values found in metal
rich \hii\ regions locally
\citep[e.g.][]{bresolinAbundancesMetalrichII2004}, where I assumed an
intrinsic $\siii{9533}/]\siii{9069}=2.58$ using atomic parameters from
NIST. While no very firm conclusion can be made from this alone, the
most likely conclusion appears to be that there is a contribution of
non-star formation activity in this source to boost the \nii{6584}
flux.

Of the other galaxies, 9239 has \sii{6717,6731}/\ha$=0.22$ which
places it in the middle of the distribution of this ratio for
star-forming galaxies
\citep[e.g.][]{2002ApJS..142...35K,2008MNRAS.385..769B}, and its
\siii{9533}/\ha\ ratio is also consistent with star formation.

Turning now to the more interesting \feii{} lines, the only regularly
detected line is \feii{1.257} which we find in 3042, 9239, and
9483. This line is seen in nearby starbursts
\citep[e.g.][]{vanziIntegralFieldNearinfrared2008,cresciIntegralfieldNearinfraredSpectroscopy2010,izotovNearinfraredSpectroscopyLarge2016a}
but it is typically quite weak at a few percent of \hb. The exception
is in shock dominated regions or in the narrow-line regions of AGN
\citep[e.g.][]{vanderwerfNearInfraredLineImaging1993,alonso-herreroUsingNearInfraredFe1997,mouriExcitationMechanismNearInfrared2000,rosenbergFeIITracerSupernova2012a}
where it can reach much higher values.

\begin{figure}
  \centering
  \includegraphics[width=\columnwidth]{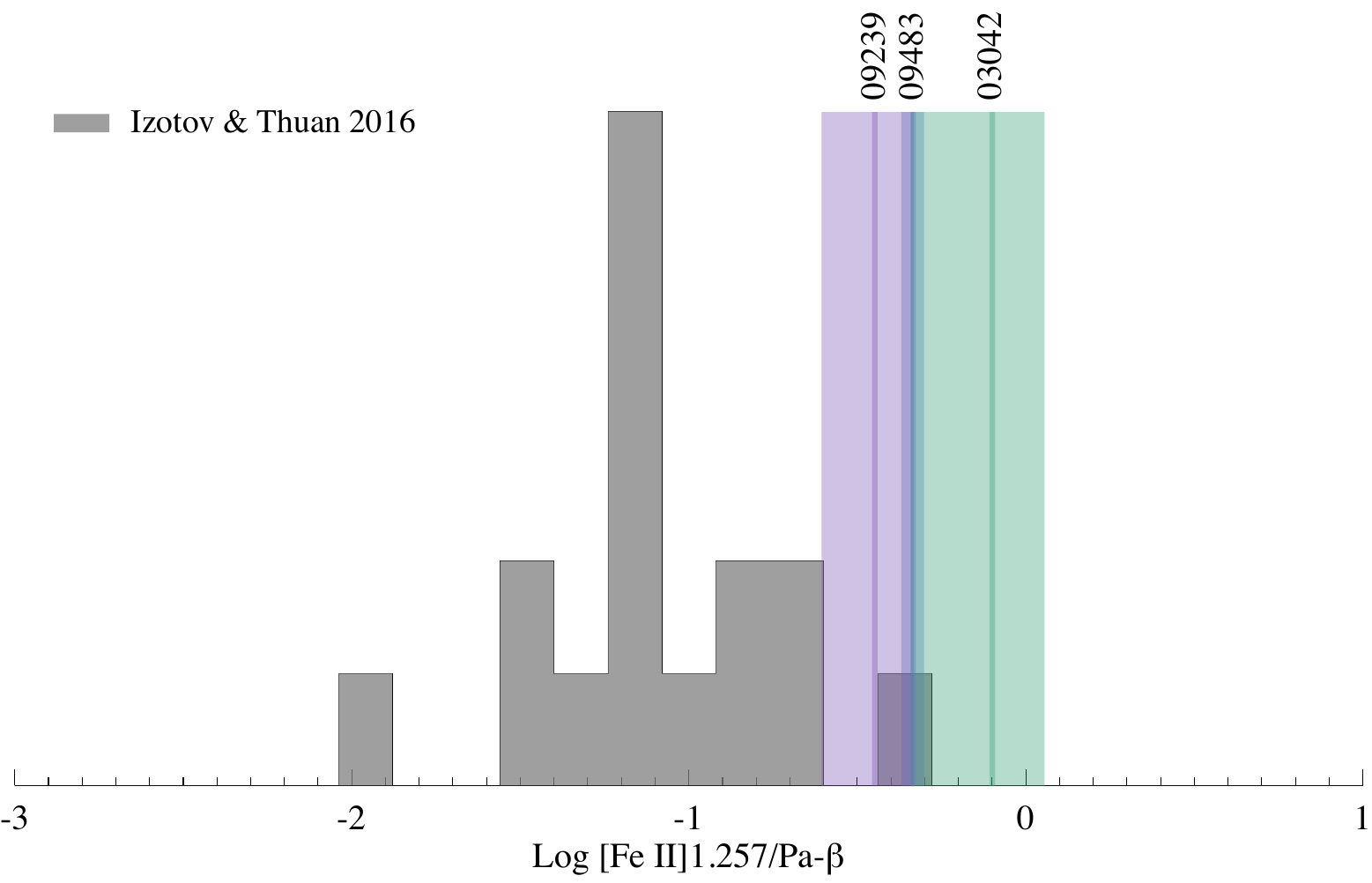}
  \caption{The ratio of \feii{1.257} to \Pb\ compared to the values
    found for low redshift BCDs
    by~\citet{izotovNearinfraredSpectroscopyLarge2016a} in gray.}
  \label{fig:it2016_comparison}
\end{figure}

\begin{figure}
  \centering
  \includegraphics[width=\columnwidth]{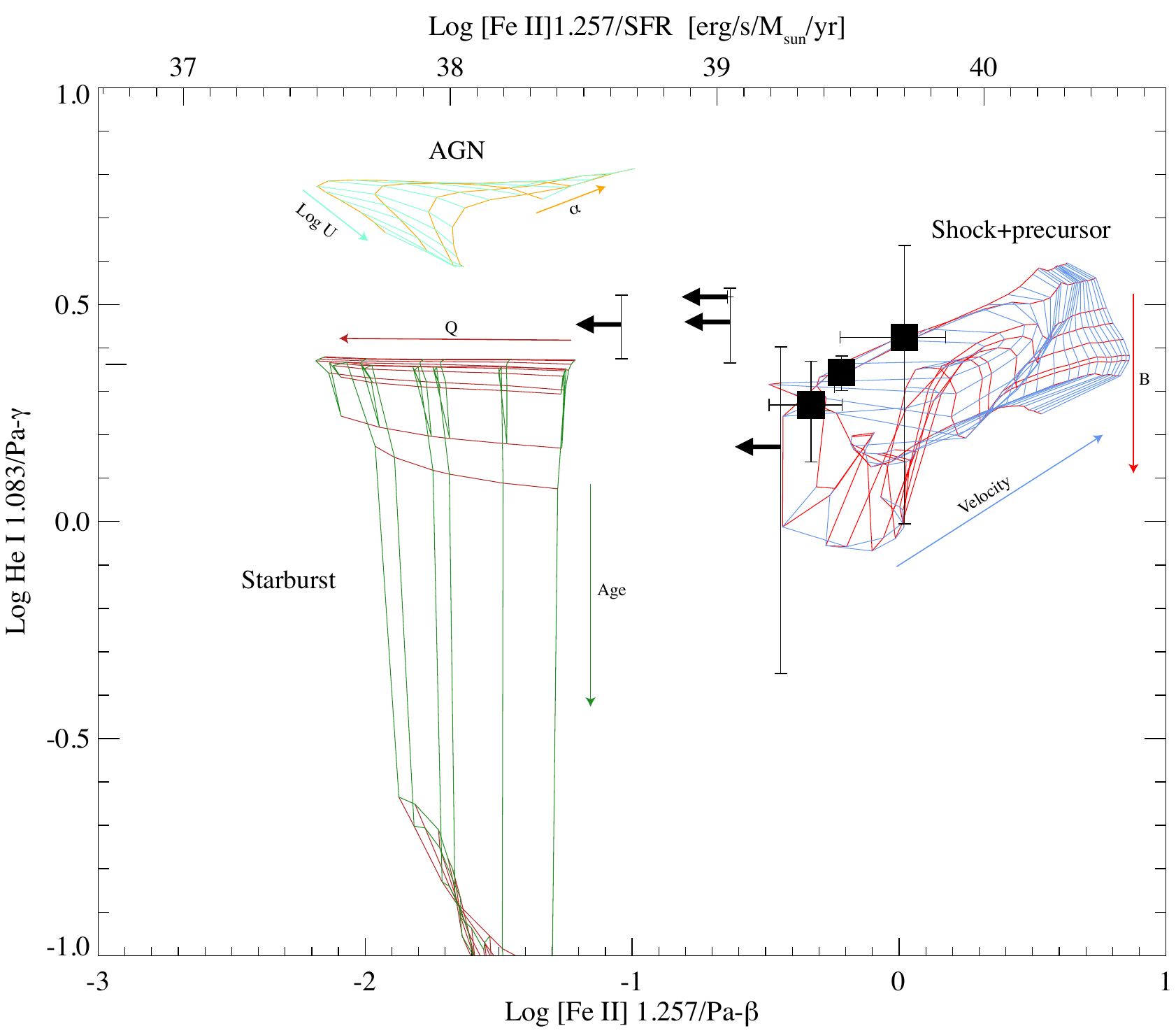}
  \caption{A diagnostic diagram of log \feii{1.257}/\Pb\ versus
    \hei{1.083}/\Pg. This is an extinction and flux-calibration robust
    diagram and the data for 3042, 9239, and 9483 are shown as the
    solid squares. The other five galaxies are shown as upper
    limits. This is contrasted to three model grids from the
    literature. The grid labeled Starburst is taken
    from~\citet{levesqueTheoreticalModelingStarForming2010}, the one
    labelled AGN
    from~\citet{grovesDustyRadiationPressureDominated2004,grovesDustyRadiationPressure2004},
    and the Shock models from~\citet{allenMAPPINGSIIILibrary2008}, see
    the text for details on the models. The separation between the
    different models is clear and is a generic feature. All three
    galaxies with detected \feii{1.257} emission are clearly most
    consistent with slow, magnetic shock models.}
  \label{fig:feii_hei}
\end{figure}

Indeed, the \feii{} lines seen in these spectra are all much too
bright to be caused by star formation, and much brighter than normally
seen in nearby starburst galaxies. To demonstrate this,
Figure~\ref{fig:it2016_comparison} compares the measured
\feii{1.257}/\Pb\ ratios in the three galaxies against the ratios
measured in 24 nearby BCDs and \hii\ regions
by~\citet{izotovNearinfraredSpectroscopyLarge2016a}. It is clear that
with the exception of the \hii\ region J1038+5330 in NGC 3310 noted by
Izotov \& Thuan, all the low-z galaxies have much weaker \feii{}
emission.

We can contrast this to models more directly because \feii{1.257} is
fairly close to \Pb\ and if I combine this with \hei{1.083}/\Pg\ which
again is a ratio of nearby lines, I get figure~\ref{fig:feii_hei}.

This figure shows the data for the three galaxies with detected
\feii{1.257} as the solid squares with error-bars. The other galaxies
are shown at the $1\sigma$ upper limit for the \feii{1.257} flux. I
contrast this against three different model libraries. I have taken a
single example grid from three libraries in the literature: the
high-massloss starburst grid with $Z/Z_\odot=0.2$ from
\citet{levesqueTheoreticalModelingStarForming2010}, a solar
metallicity $n=10^4\,\mathrm{cm}^{-3}$ dusty AGN model grid
from~\citet{grovesDustyRadiationPressureDominated2004,grovesDustyRadiationPressure2004},
and a solar metallicity $n=0.01\mathrm{cm}^{-3}$ shock model grid
from~\citet{allenMAPPINGSIIILibrary2008} where I used the
shock+precursor grid. All models were obtained through the
\texttt{itera} code by~\citet{grovesITERAIDLTool2010a}. 

The immediate observation we can make is that the three SMACS galaxies
with clearly detected \feii{1.257} all fall clearly in the shock
dominated region. They lie towards the high $B$-field, low velocity
region of the shown grid. In reality these galaxies are likely to have
a mix of shocks, star-formation and AGN contributing to their emission
lines and a more detailed modeling would be required to disentangle
these and the present data are not of sufficient quality to warrant
this. Furthermore, for some of these galaxies the aperture
  corrections are substantial, this is shown in
  Table~\ref{tab:zlowzlum} which provides line luminosities on spectra
  normalised to aperture fluxes and total fluxes. That said, however, it will
be of considerable interest to understand these ``shock-dominated''
galaxies in more detail and to understand how they fit into the
  overall population of galaxies at these redshifts \citep[see also][]{reddyPaschenlineConstraintsDust2023}. It would also be very desirable to have
additional diagnostic ratios to further understand the nature of these
sources, but only 3042 has some qualitatively different lines, in
\nii{6583}, \siii{9533}, and \ha\, and those are not strongly
diagnostic of shock activity
\citep{allenMAPPINGSIIILibrary2008}. Instead~\citet{allenMAPPINGSIIILibrary2008}
recommended UV and UV-optical diagnostics such as
\ciii{1909}/\cii{2326} and \neiii{3869}/\nev{3426}, neither of which
are available for our galaxies but which conceivably could be obtained
for $z\sim 1$ galaxies with deep ground-based observations.

The final set of lines of key interest are the \hmol\ lines. These
originate in the cooler medium outside the \hii\ regions and as such
provide very complementary information to lines coming from hotter
regions of the ISM. In the present sample only 9483 shows strong
\hmol\ lines and even then only three: \hmol\ 1-0 S(3), \hmol\ 1-0 S(2)
and \hmol\ 1-0 S(1). \hmol\ 1-0 S(3) is contaminated by \hei{1.955},
leaving only the \hmol\ 1-0 S(2)/\hmol\ 1-0 S(1) ratio as a
diagnostic. This ratio has a value of $0.68\pm 0.47$ which is
comparable to the values found
by~\citet{izotovNearinfraredSpectroscopyLarge2016a} for nearby
star-forming galaxies and regions (0.3--0.8) and the value of 0.28
found for the Orion Bar
by~\citet{kaplanExcitationMolecularHydrogen2017}. Since we saw above
that 9483 has a significant contribution of shocks to its \feii{1.257}
line flux, one might expect that the \hmol\ line ratios should
approach that corresponding to a thermal distribution. Unfortunately
the detected lines are not very diagnostic for this as the models for
thermal and fluorescent emission
in~\citet{blackFluorescentExcitationInterstellar1987} span this range,
with a preference for a lower value for thermal distributions and
somewhat higher for fluorescent emission but in either case consistent
with the observed line ratio. Future observations will surely provide
much more information on this and open up the diagnostic potential of
\hmol\ lines for galaxies out to $z\sim 1.3$ with NIRSpec.

\section{Comparison to local galaxies}
\label{sec:comp_to_local}

As discussed in the introduction, analogues of high-z galaxies in the
local Universe have been sought after for a long time, and naturally
the first papers looking at these NIRSpec data have discussed
extensively their properties in the context of nearby
galaxies. Thus~\citet{schaererFirstLookJWST2022} compared the sample
of extreme line-emitters from~\citet{izotovMultiwavelengthStudy142014}
and galaxies from the Low-Z Lyman Continuum Survey \citep[LzLCS,
][]{fluryLowredshiftLymanContinuum2022} to the NIRSpec sample, finding
considerable correspondences. \citet{rhoadsFindingPeasEarly2023}
compared the NIRSpec sample to nearby Green Pea galaxies
\citep[e.g.][]{cardamoneGalaxyZooGreen2009a,jaskotOriginOpticalDepth2013},
specifically from the sample defined
in~\citet{jiangDirectMetallicityCalibration2019}
and~\citet{yangLyaGalaxiesEpoch2019}. A mixture of those approaches
was taken by~\citet{trumpPhysicalConditionsEmissionLine2023} who
compared both to Green Peas from
\citet{brunkerPropertiesKISSGreen2020} but also to extreme line
emitters from~\citet{perez-monteroExtremeEmissionlineGalaxies2021a}. A
similar approach was taken by~\citet{katzFirstInsightsISM2023} who
compared against compact galaxies of various colour
\citep{yangLyaProfileDust2017,yangBlueberryGalaxiesLowest2017}, as
well as low-metallicity galaxies
\citep{izotovLowredshiftLowestmetallicityStarforming2019} and exteme
line-emitters \citep{amorinExtremeEmissionlineGalaxies2015}.  With
small variations, all these authors and comparisons find that there
are galaxies in the nearby Universe that are broadly similar to the
NIRSpec sample, although exactly how similar is open to some
discussion.

What is noticeable is that these studies all limit themselves to star
forming galaxies, thus there is an \textit{a priori} assumption that
the ionization source in these galaxies must be star formation. This
might very well be correct and it is not an unreasonable assumption,
but as we saw above, 6355 appears to have an AGN contribution and it
is notoriously difficult to distinguish between AGN and star formation
as ionization source at low metallicity. Thus here I take a slightly
different approach. I will not start with a particular sample but
rather I will ask: what galaxies in the local Universe have similar
emission line properties to the NIRSpec sample and what are the
properties of this local sample?

Since I am interested in excitation properties, I will focus on line
ratios and ignore extrinsic quantities like size or mass. Specifically
I always require the \oiii{5007}/\hb\ ratio to match the NIRSpec
sample within 1$\sigma$, and then define two samples: the
\oiii{4363}/\hg\ counterparts and the \neiii{3869}/\oii{3727}
counterparts. I will base myself on the SDSS DR7 sample used in B13
and only use galaxies with S/N$>7$ in all relevant lines. With these
preambles we find the following results:
\begin{itemize}
  \item \textbf{4590}: there are 13 galaxies with \oiii{4363}/\hg\
  within 1$\sigma$. 6 of these are AGN, while 4 are star-forming, 2
  composite and one unclassified. The \neiii{3869}/\oii{3727} ratio
  gives a similar result with 2/3 AGNs and 1/3 star-forming.
  \item \textbf{5144}: there are 192 galaxies with \oiii{4363}/\hg\ within
  1$\sigma$ of 5144. 73\% of these are star-forming and all but 2 of
  the remainder are classified as AGN. The \neiii{3869}/\oii{3727}
  ratio is matched by no galaxies in the SDSS within 1$\sigma$.
  \item \textbf{6355}: there are 155 galaxies that are close to the
  \oiii{4363}/\hg\ ratio. Of these 76\% are star-forming while the
  remainder are almost all AGN. When considering
  \neiii{3869}/\oii{3727} only 40 galaxies are within 1$\sigma$ of the
  NIRSpec data. Out of these 55\% are AGN, 2 unclassified, and the
  rest star-forming.
  \item \textbf{8140}: as I do not consider this to have \oiii{4363} I
  do not consider this ratio, but for the \neiii{3869}/\oii{3727}
  ratio a total of 371 galaxies in the SDSS fall within
  1$\sigma$. These are composed of 15\% star-forming and 82\% AGN.
  \item \textbf{10612}: there are 172 SDSS galaxies with
  \oiii{4363}/\hg\ within 1$\sigma$ of the NIRSpec data. Out of these
  only 3 are classed as star-forming, with the remainder all falling in
  the AGN part of the BPT diagram. There are only two galaxies that
  have \neiii{3869}/\oii{3727} within 1$\sigma$ of the NIRSpec data,
  one AGN, one star-forming galaxy. 
\end{itemize}

\begin{figure}
  \centering
  \includegraphics[width=\columnwidth]{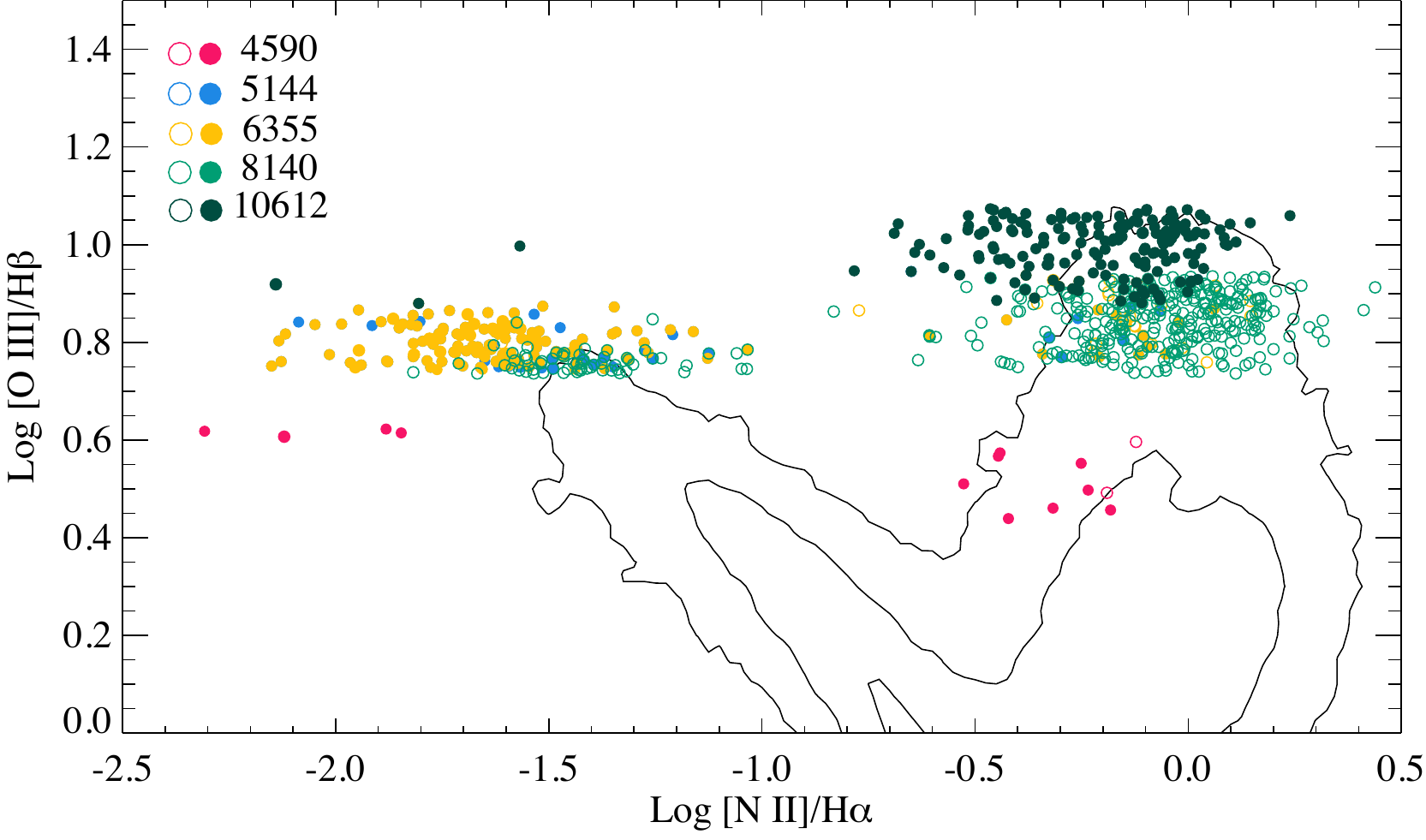}
  \caption{The location of analogues to the high-z NIRSpec galaxies in
    the \nii{6584}/\ha\ versus \oiii{5007}/\hb\ diagram. The solid
    contours show the distribution of the main sample of low-z
    galaxies from the SDSS DR7. The filled symbols show
    \oiii{4363}/\hg\ analogues while the open symbols refer to
    \neiii{3869}/\oii{3727} analogues.}
  \label{fig:bpt_analogues}
\end{figure}

These results are summarised in Figure~\ref{fig:bpt_analogues} where
filled disks show the \oiii{4363}/\hg\ analogues while the open
symbols the \neiii{3869}/\oii{3727} ones. It is evident that based on
the observed line ratios the local counter-parts will fall in two
regions: they are either low-Z star-forming galaxies (or low-Z AGNs we
should not forget), or clear AGN. Taken at face value this suggests
that we should at least take the assumption that all these are powered
by star formation with a grain of salt, although one should not
over-interpret the results of the local sample either. In particular
the 10612 which is easily the most extreme galaxy in the sample, must
be questioned as to its ionization source as also pointed out
by~\citet{schaererFirstLookJWST2022}.

For the \oiii{4363}/\hg\ analogues I have also co-added the PDFs of
all their derived parameters from the fits described
in~\citet{brinchmannEstimatingGasMasses2013}. In the following section
(see also Figure~\ref{fig:cl01_G16_OH_results}), I will compare the
resulting distributions to the high-z galaxies. This is only possible
for star-forming galaxies, thus when reading the following section it
is important to keep in mind that there is a possibility that several
of these sources have a contribution of an AGN to their line fluxes
which will make some of the following results less conclusive.

\section{Gas-phase properties of the galaxies}
\label{sec:gas_phase}

\begin{table*}
\renewcommand{\arraystretch}{1.4}
\caption{Results from the CL01 and G16 photoionization fitting. The
  central value is the median of the PDF and the uncertainties quoted
  are 16\% and 84\% of the PDFs for the indicated quantity and do not
  include systematic uncertainties. The depletion time assumes an
  effective area probed by emission lines of $1\mathrm{kpc}^2$.}
 \label{tab:pifit}
 \begin{tabular}{lrrrrr}\hline
Quantity & \multicolumn{1}{c}{ 4590}  & \multicolumn{1}{c}{ 5144}  &
                                                                     \multicolumn{1}{c}{ 6355}  & \multicolumn{1}{c}{ 8140}  & \multicolumn{1}{c}{10612}   \\ \hline
$12+\log \mathrm{O/H}$ (CL01) &  $7.60^{+0.04}_{-0.04}$ &  $7.89^{+0.26}_{-0.13}$ &  $8.01^{+0.11}_{-0.10}$ &  $8.44^{+0.10}_{-0.10}$ &  $7.85^{+0.09}_{-0.08}$ \\
$12+\log \mathrm{O/H}$ (G16) &  $7.35^{+0.04}_{-0.05}$ &  $7.63^{+1.26}_{-0.20}$ &  $8.28^{+0.37}_{-0.21}$ &  $8.44^{+0.09}_{-0.08}$ &  $7.91^{+0.11}_{-0.09}$ \\
Log U (CL01) &  $-2.03^{+0.02}_{-0.02}$ &  $-2.65^{+0.18}_{-0.14}$ &  $-2.27^{+0.06}_{-0.05}$ &  $-2.74^{+0.11}_{-0.06}$ &  $-2.03^{+0.02}_{-0.02}$ \\
Log U (G16) &  $-1.63^{+0.21}_{-0.15}$ &  $-2.35^{+0.53}_{-0.30}$ &  $-1.97^{+0.24}_{-0.15}$ &  $-2.28^{+0.08}_{-0.06}$ &  $-1.08^{+0.05}_{-0.11}$ \\
$\Sigma_{\mathrm{gas}} [\mathrm{M}_\odot \mathrm{pc}^{-2}]$ (CL01) &  $2.20^{+0.04}_{-0.04}$ &  $1.93^{+0.40}_{-0.47}$ &  $1.97^{+0.38}_{-0.41}$ &  $0.59^{+0.16}_{-0.17}$ &  $2.25^{+0.36}_{-0.24}$ \\
$\Sigma_{\mathrm{gas}} [\mathrm{M}_\odot \mathrm{pc}^{-2}]$ (G16)) &  $2.86^{+0.31}_{-0.24}$ &  $2.02^{+0.61}_{-0.84}$ &  $1.58^{+0.37}_{-0.53}$ &  $0.36^{+0.34}_{-0.12}$ &  $2.19^{+0.31}_{-0.38}$ \\
Log SFR (CL01) &  $0.88^{+0.02}_{-0.02}$ &  $0.57^{+0.13}_{-0.11}$ &  $1.34^{+0.12}_{-0.16}$ &  $0.21^{+0.05}_{-0.04}$ &  $0.99^{+0.04}_{-0.06}$ \\
Log SFR (G16) &  $0.91^{+0.11}_{-0.12}$ &  $0.50^{+0.65}_{-0.30}$ &  $1.29^{+0.39}_{-0.20}$ &  $0.17^{+0.16}_{-0.08}$ &  $1.23^{+0.29}_{-0.26}$ \\
$\log \Sigma_{\mathrm{gas}}/\Sigma_{\mathrm{SFR}}$ [yrs] (CL01) &  $6.10^{+0.36}_{-0.36}$ &  $6.89^{+0.38}_{-0.24}$ &  $6.73^{+0.18}_{-0.18}$ &  $6.54^{+0.07}_{-0.07}$ &  $7.02^{+0.39}_{-0.26}$ \\
 \end{tabular}
\end{table*}

\begin{figure*}
  \centering
  \includegraphics[width=0.8\textwidth]{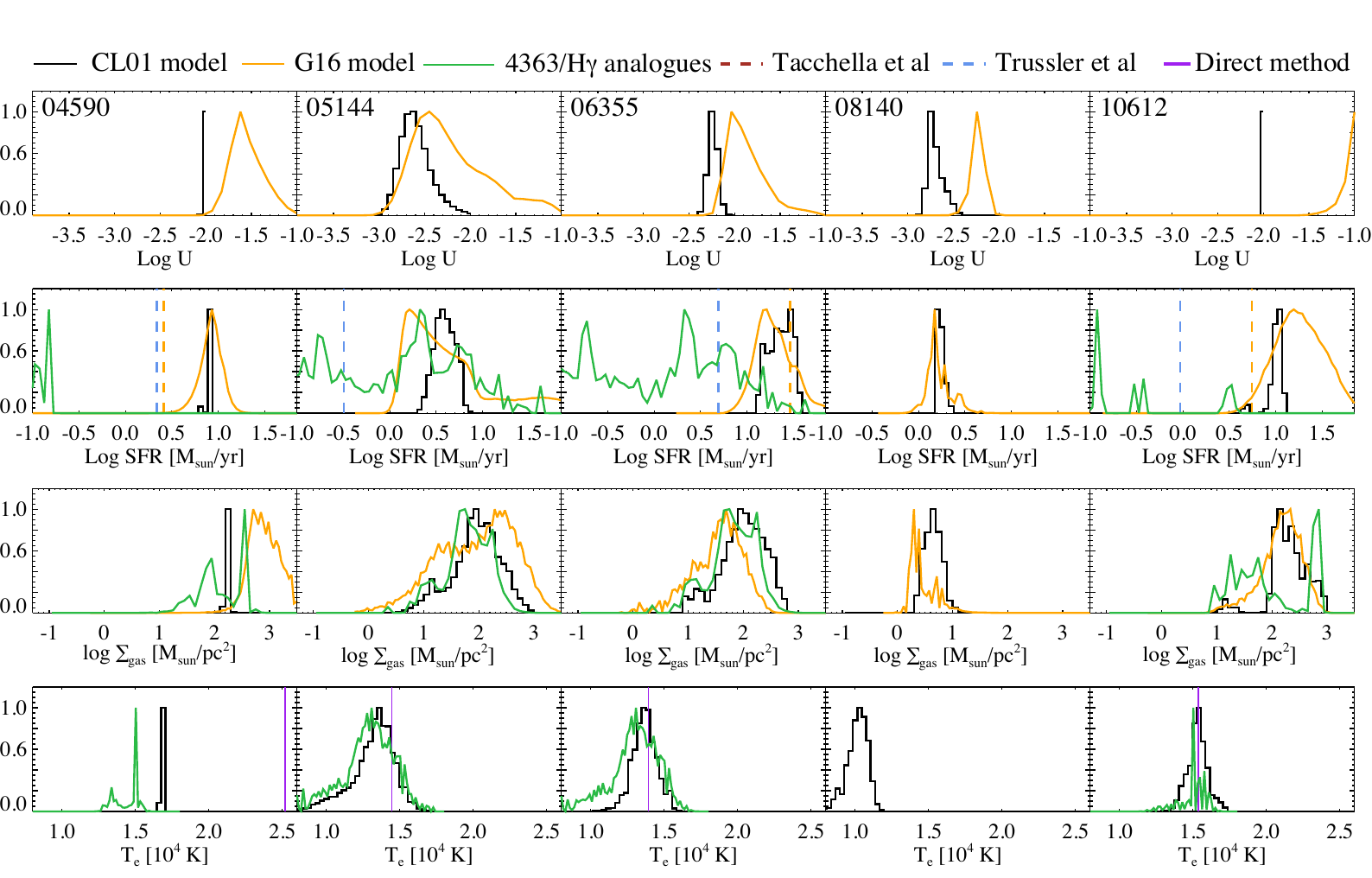}
  \caption{The results of fitting CL01 models to the five
    high-redshift galaxies. Each row shows the probability
    distribution function (PDF) of a single parameter. The top row
    shows the ionization parameter, the second the logarithm of the
    star formation rate, the third row the surface mass density of
    gas, and the final row the estimated electron tempearture of the
    gas. The PDFs have not been closed at the parameter grid edge
    which is clearly seen for $\log U$ for 5144, 6355, and in
    particular 10612. The fit results for these three galaxies are
    therefore uncertain. The G16 model is shown in orange in the
    panels where predictions could be obtained. In the SFR row I
    compare the results to the photometrically estimated SFR
    from~\citet{trusslerSeeingSharperDeeper2022}, and the
    spectrum+photometry determination
    from~\citet{tacchellaJWSTNIRCamNIRSpec2023}, and in the $T_e$ row
    I show the $T_e($\oiii{}$)$ values from the direct method as
    purple vertical lines. The green lines show the distributions for
    the \oiii{4363}/\hg\ analogues defined in
    section~\ref{sec:comp_to_local}.}
  \label{fig:cl01_results}
\end{figure*}

Returning now to the CL01 and G16 photoionization model fits to the
data, Figure~\ref{fig:cl01_results} shows multiple fit parameters for
the five high-$z$ galaxies. The first row shows the ionization
parameter for each galaxy and here we immediately see one of the
limitations of the CL01 grid: it does not go to sufficiently high
ionization parameters and the PDFs are pushed up towards the edge. For
those galaxies for which this is noticeable, 5144 and 10612 in
particular, this puts the other results into question although we saw
in Figure~\ref{fig:triangle_G16} that $\log U$ is relatively
independent of other parameters in the fit. I do not show the analogue
sample in this row as the CL01 model is such a poor match in this
respect.

The second row shows the star formation rate, derived from the scaling
parameter in equation~\eqref{eq:Pmodel}. The star formation rates
inferred from the emission lines are corrected for lensing using the
same values as~\citet{trusslerSeeingSharperDeeper2022} which are taken
from~\citet{pascaleUnscramblingLensedGalaxies2022}. The dashed blue
lines show the Star Formation Rate (SFR) estimates from the photometry
presented in~\citet{trusslerSeeingSharperDeeper2022} (who did not
provide a value for 8140) and the dashed burgundy lines those
presented by~\citet{tacchellaJWSTNIRCamNIRSpec2023} adjusted to use
the same magnification values
as~\citet{trusslerSeeingSharperDeeper2022}. They all agree to within a
factor of 3, but often better --- indeed the agreement
with~\citet{tacchellaJWSTNIRCamNIRSpec2023} is considerably better
than this despite very different methodologies. However the
spectroscopic estimates should be considered to have a substantial
systematic uncertainty due to the uncertain flux calibration. The SFR
distributions for the local analogues show a very large spread which
is not surprising as the matching was done ignoring all extrinstic
quantities.

The third row shows an estimate of the surface gas mass density of the
galaxies following the methodology of B13 (see that reference for
details). The results are interesting in that they prefer high surface
mass densities for all the high SFR sources. These gas estimates are
in reasonable agreement between CL01 and G16 when the CL01 models are
adequate fits, and it is also clear that the local analogues have
similar gas properties to the $z>5$ galaxies. If we assume that the
emission lines come from a region with area $\sim 1\,\mathrm{kpc}^2$
which is not entirely out of question given the small sizes, we find
depletion times of
$t_{\mathrm{depl}} = \Sigma_{\mathrm{gas}}/\Sigma_{\mathrm{SFR}}
\approx 10^7\,\mathrm{yrs}$ (see Table~\ref{tab:pifit}). This is short
but very similar to the values found for starbursting galaxies in B13,
and it is in good agreement with the conclusion
of~\citet{tacchellaJWSTNIRCamNIRSpec2023} that these appear to be
rapidly accreting galaxies.

The final row shows electron temperature estimates from the models,
compared to those obtained from the direct method. The $T_e$ values
from the CL01 fits are averages over the Str{\"o}mgren sphere which in
general leads to lower temperatures than from \oiii{4363}/\oiii{5007}
but for both 6355 and 10612, the CL01 fits give temperatures in good
agreement with the direct method. While this could be interpreted as
evidence for very hot ionized gas where the \oiii{4363}/\oiii{5007}
provides a good estimate of the mean temperature, the fact that the
CL01 model is a poor fit for both of these argues against any strong
conclusion but it does highlight the usefulness of using
photo-ionization models to estimate electron temperatures as well.  In
all but 4590 the CL01 fit to the low-$z$ analogues also results in
temperatures similar to the high-$z$ galaxies.

These quantities, including the depletion time are all given in
Table~\ref{tab:pifit} for the five $z>5$ galaxies with sufficient
lines. 4580 does not have sufficient lines for a photoionization
model fit to be useful. It is also important to keep in mind that the
errors given in the table do not include systematic uncertainties
which given the calibration uncertainties could be substantial.

\section{Discussion}
\label{sec:discussion}

Despite the lingering calibration issues, these data are demonstrating
how ground-breaking NIRSpec will be for the study of ionized gas in
galaxies. 

The $z<3$ galaxies are showing a surprising amount of shock excited
gas signaled by an elevated \feii{1.257} emission. In the local
Universe this is not seen in low-mass star-bursts
\citep{izotovNearinfraredSpectroscopyLarge2016a} and regions where
\feii{} is significantly enhanced tends to be associated with shocked
regions or AGNs. The shocks in this case is usually assumed to be
associated with supernova remnants \citep[SNRs,
][]{greenhouseNearInfraredFeIi1991,alonso-herreroFeII6442003,rosenbergFeIITracerSupernova2012a}. \citet{bruursemaSearchSupernovaRemnants2014}
used this to search for SNRs in NGC 6946 and found a typical
luminosity per SNR of $\sim
10^{36}\,\mathrm{erg}\,\mathrm{s}^{-1}$. It should be noted that there
is a spread of almost an order of magnitude in this luminosity and
other studies find different mean values, but if we adopt
the~\citeauthor{bruursemaSearchSupernovaRemnants2014} value, we find
that the \feii{1.257} luminosity of our three sources corresponds to
2--3$\times 10^3$ \feii{}-luminous SNR. Thus, if the \feii{1.257} all
comes from SNR, we can estimate a supernova rate
$\sim 0.1\,\mathrm{yr}^{-1}$, assuming a typical life-time for the
bright radiative shocks of $\sim 10^4\,\mathrm{yr}$
\citep{vinkPhysicsEvolutionSupernova2020}. This is not an unreasonable
rate of star formation, thus this presents a plausible scenario for
the brightness of the \feii{} lines. However it does not address the
question of why other galaxies (1917, 5735, 8506, 9721, 9922) show no
sign of \feii{1.257} despite also having bright Paschen lines.  Indeed
in figure~\ref{fig:feii_hei} the top $x$-axis shows the \feii{1.257}
luminosity per SFR, and as can be seen, this varies strongly between
the galaxies

Turning now to the high redshift galaixes, several of the papers on
these galaxies have commented on the high value of \oiii{4363}/\hg\ or
\oiii{4363}/\oiii{5007} in 4590. I have avoided this until now because
I do not feel there is much reason to worry to much about
this. Firstly, the \oiii{4363} line is weak, only detected at
3--4$\sigma$ depending on the noise estimate adopted. Secondly, the other
lines in the galaxy do not show particularly extreme properties thus
there is no clear reason to think that this is anything but a
statistical fluke. Indeed the fits to the G16 model above predicts a
\oiii{4363} flux somewhat lower than that observed which would make
the galaxy much more normal.

In contrast, 6355 appears to show \neiv{2422,2424} in emission and is
most likely a narrow-line AGN, while 10612 has an ionization parameter
$\log U>-1$ which is normally the regime occupied by AGNs. One might
object that \heii{4686} is not seen, but the expected flux given the
strong-line fluxes is below $10^{-21}\,$erg/s/cm$^{-2}$ (the G16 model
predicts $2\times 10^{-22}\,$erg/s/cm$^{-2}$ for an extreme
star-forming model) which is below the detection threshold of the
data. This then argues for potentially at least two out of the five
high-$z$ galaxies being affected by AGN activity.

It is also clear that the low-$z$ analogues are showing intrinsic
properties very well matched to the high-$z$ galaxies and this appears
to justify the long-standing quest to carefully characterise these to
contrast to high-$z$ galaxies. It also allows us to examine a
potential bias that might affect all NIRSpec studies of high-$z$
galaxies.

\subsection{Selection of galaxies and biases in the mass-metallicity
  relation}
\label{sec:mass-met-bias}

Several papers
\citep[e.g.][]{schaererFirstLookJWST2022,curtiChemicalEnrichmentEarly2023}
have already tried to assess the evolution of the mass-metallicity
(MZ) relation with redshift by comparing the JWST SMACS results to
$z\sim 0$
\citep{tremontiOriginMassMetallicity2004,yatesRelationMetallicityStellar2012,andrewsMassMetallicityRelationDirect2013,telfordExploringSystematicEffects2016,curtiMassMetallicityFundamental2020}
and $z\sim 2$ results
\citep[e.g.][]{sandersMOSDEFSurveyDirectmethod2020}.  With only three
galaxies, selected in a somewhat haphazard manner and still
substantial uncertainties in the stellar mass estimates
\citep[e.g.][]{schaererFirstLookJWST2022,tacchellaJWSTNIRCamNIRSpec2023,carnallFirstLookSMACS07232023}
this is of course preliminary. That aside, these studies do
demonstrate the potential of future, larger, JWST studies to make real
progress here.

\begin{figure*}
  \centering
  \includegraphics[width=\textwidth]{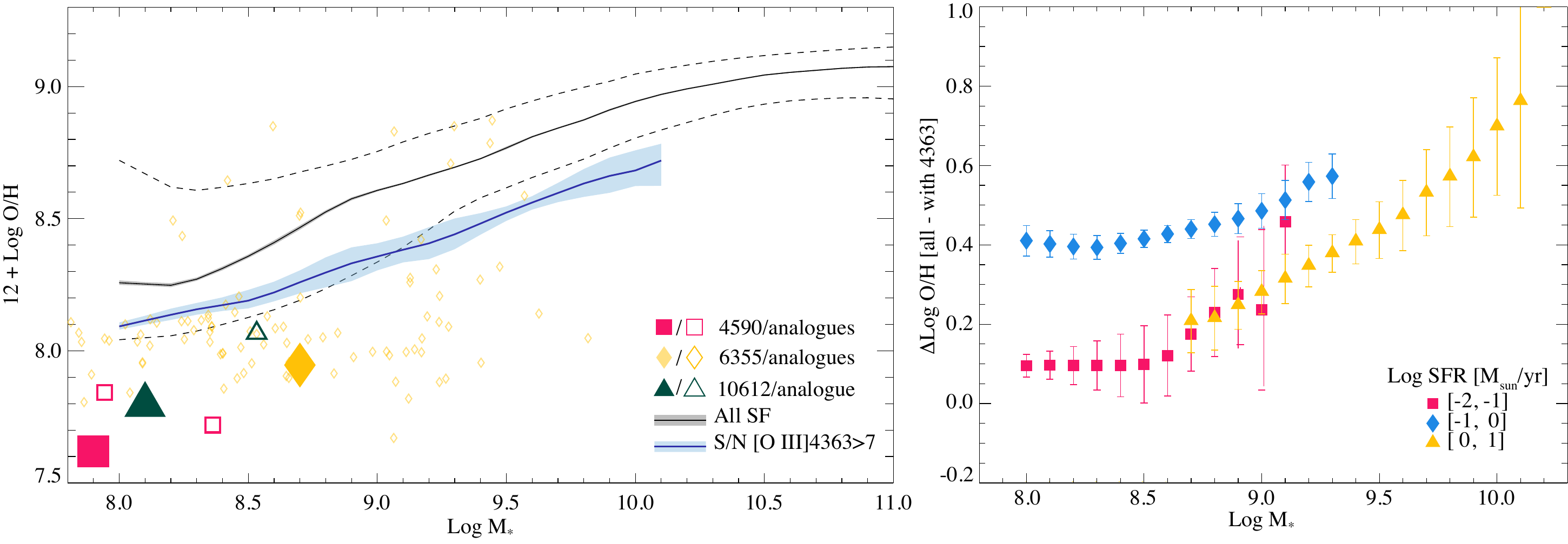}
  \caption{Left: The mass-metallicity relation for star-forming
    galaxies in the SDSS DR7, compared to different selection methods
    and the three highest redshift galaxies from the current JWST
    sample. The solid black line shows the median trend for all
    galaxies with the dashed lines showing the 68\% spread around
    this. Blue solid line: the median trendfor star-forming galaxies
    with \oiii{4363} detected with a S/N$>7$. The filled symbols show
    the locations of the high-$z$ galaxies (square: 4590, diamond:
    6355, triangle: 10612) using masses
    from~\citet{tacchellaJWSTNIRCamNIRSpec2023} and CL01-based
    metallicities for a direct comparison to SDSS. The corresponding
    open symbols shows the \oiii{4363}/\hg\ analogues for each
    high-$z$ galaxy. Right panel: the difference in the median
    $12+\log \mathrm{O/H}$ between all galaxies and those with
    $\mathrm{S/N}>7$ in \oiii{4363} as a function of mass and star
    formation rate.}
  \label{fig:MZ-bias}
\end{figure*}

There are many potential biases when constructing the MZ relation and
this has been discussed in the literature
\citep[see][]{telfordExploringSystematicEffects2016,cresciFundamentalMetallicityRelation2019},
but one important potential bias for the current scenario is the
requirement to have a detected \oiii{4363}. It is possible to make recourse
to the local sample again to understand how this comes about. Note
that the bias I discuss here also applies to the
mass-metallicity-SFR
\citep{2010MNRAS.408.2115M,lara-lopezFundamentalPlaneField2010}.

Since \oiii{4363} is exponentially sensitive to the temperature of the
ionized gas, requiring the detection of this line means that you will
always get a sample of the lowest metallicity galaxies at each
mass. This bias was already commented on indirectly
by~\citet{sandersMOSDEFSurveyDirectmethod2020} who focused their
discussion on the construction of calibration relations and the SFR
distributions of the samples used to construct these. The bias is
however more directly affecting the MZ relationship as it becomes more
pronounced as one moves to higher masses (higher mean metallicities). 

To illustrate this, the left panel of Figure~\ref{fig:MZ-bias} shows
the MZ plane for the SDSS DR7 and the effect of selection on
\oiii{4363} detectability. The black line shows the median trend for
all SF galaxies from the SDSS DR7 sample of B13 and the dashed lines
the 68\% spread around this with the grey shading showing the 68\%
uncertainty on the median as determined from bootstrap resampling
including Monte Carlo sampling of the uncertainties on $\log M_*$ and
$12+\log \mathrm{O/H}$. To reduce aperture correction effects, I
have limited my sample to have $z>0.01$ and a half-light diameter no
more than 3 times the fibre diameter of the SDSS, but I have not
removed failed deblends as done by e.g.\
\citet{andrewsMassMetallicityRelationDirect2013}; the conclusions here
are not significantly affected by the details of these criteria. 

The dark blue line with light blue shading shows the same, but now
requiring a $\mathrm{S/N}>7$ in \oiii{4363} with the shading
indicating the 68\% uncertainty on the median. It is clear that
applying this criterion leads to a median trend that is offset by
0.1--0.25 dex and with a flatter slope than the main sample. Thus one
would conclude that this sample is significantly offset --- despite
using the same stellar mass and metallicity indicators here.

One might rightly object that the SFR distribution of the two samples
are different, so in the right panel I show the difference between the
main sample and the sample requiring a $\mathrm{S/N}>7$ in \oiii{4363}
as a function of mass divided in three bins of $\log \mathrm{SFR}$ as
measured within the fibre, as indicated in the legend. The errorbars
correspond to uncertainties on the median, the 68\% spread is much
larger. It is clear that there is a systematic trend with mass and it
is even larger than when not controlling for SFR. The differences
  between the different SFR bins are small, but systematic: at low SFR
  the sample of all versus that with \oiii{4363} are nearly coincident
  so very little difference is seen. At somewhat higher SFRs we have
  the largest difference in sample and at very high SFR and low $M_*$
  we again have very large overlap between the samples. One should,
however, be cautious with overinterpreting this figure, however, as
the SDSS is not complete below $\log M_*/M_\odot \sim 9$ and the
sample of galaxies with \oiii{4363} does include galaxies that were
not targetted as such. However the basic lesson appears robust:
constructing MZ relations based on direct method metallicities
requires very careful assessment of selection biases when comparing
samples.

To put this in the context of the high-$z$ galaxies, I have also
inserted these (filled symbols) with their \oiii{4363}/\hg\
star-forming analogues (open symbols) in the figure. The squares show
4590 and its analogues, the diamonds 6355 and its analogues and the
triangles 10612 and its sole star-forming analogues. For this plot I
have used stellar masses from~\citet{tacchellaJWSTNIRCamNIRSpec2023}
and the CL01 model to estimate metallicities in order to compare
directly to SDSS DR7, however a very similar picture is found using
the G16 model.

Does this mean that this bias will also apply at high redshift? This
is one of the questions that upcoming JWST surveys with NIRSpec will
be able to tell us --- it depends on the spread in metallicity at a
given mass. If this is very small at high redshift, then a sample
defined to have detected \oiii{4363} will provide a good measure of
the mean metallicity, but if there is substantial scatter then direct
method metallicities might result in a significantly biased MZ
relation, including most likely an incorrect slope.

Since we saw above that photoionization modelling gives oxygen
abundances in good agreement with the direct method, as long as the
models span a sufficiently large range in physical parameters, a
viable way to combat these potential biases is to use photoinization
models to estimate oxygen abundances for all galaxies instead --- using
\oiii{4363} to validate the results on a case-by-case basis.

\section{Conclusions}

In the preceding I have presented three new redshifts, all at $z<3$,
bringing the total number of secure redshifts for these NIRSpec
observations in the SMACS 0723 field to 14. I highlighted the
usefulness of modifying the methodology for direct temperature
estimates to use the double line ratio
\oiii{4363}/\hg/\oiii{5007}/\hb. Doing so reduces senstivity to flux
calibration or dust attenuation uncertainties by a factor of $\sim
3$. If in addition the ionic abundance determinations are referenced
to nearby hydrogen lines, one can strongly reduce effect of flux
calibration uncertainties or unknown reddening on oxygen abundance
determinations. Given the unavoidable wavelength-dependent slit losses
with NIRSpec this modification of the methodology is likely to be
beneficial for future high-$z$ abundance studies.

I also find that 6355 shows a fairly clear \neiv{2422,2424} line which
I argue is evidence for this being a narrow-line AGN. The existence of
one such high-z, low-mass AGN is interesting but larger samples are
necessary before any implications for the process of reionization
become clear. However it is notable that 10612 is showing line ratios
that at low-$z$ are commonly seen in AGN, which coupled to an
ionization parameter that reaches the model boundary at $\log U=-1$,
suggests that this galaxy also might harbour an AGN, thus one must
interpret these results with some caution.

The lower redshift galaxies also seem to be rather active with three
galaxies showing very strong \feii{1.257}/\Pg\ relative to what
star-forming galaxies at $z\sim 0$ is showing. These galaxies have
emission spectra that can be explained nearly fully by shock
models and one possible explanation is that they have a particularly
high number of supernova remnants. Further data are needed to fully
understand why some galaxies show very strong \feii{} lines and others
not, while having comparable Paschen lines, but clearly aperture
corrections will play an important role here.

I also presented modelling of the high-$z$ galaxies using
photoionization models and show that these models result in inferred
oxygen abundances in good agreement with those found using the direct
method. Thus Bayesian modeling using photoionization model grids is a
viable way to estimate the physcial conditions in $z>5$ galaxies also
in conditions where \oiii{4363} is not detected, thus making for a
better technique for large samples, reducing complex selection
effects.

Finally, I showed that selecting samples based on the presence of
\oiii{4363} has the potential to significantly bias determinations of
the mass-metallicity relation and that a careful assessment of
selection effects must be made in this case and I argued that it might
be better to base oneself on Bayesian photoionization modeling of the
strong lines than to use temperature sensitive lines as the main
oxygen abundance estimation technique when doing statistical studies
such as of the mass-metallicity relation.

\section*{Acknowledgements}

Firstly, I would like to thank the JWST commissioning team for
providing these data early and giving us all a taster of what JWST
will bring in the coming years. I thank Michael Maseda, Themiya
Nanayakkara, Madusha Gunawardhana, Leindert Boogaard, Roberto Maiolino
for very helpful comments on an earlier version of the text. The
research presented here was supported by Fundação para a Ciência e a
Tecnologia (FCT) through research grants UIDB/04434/2020 and
UIDP/04434/2020. JB acknowledges financial support from the Fundação
para a Ciência e a Tecnologia (FCT) through national funds
PTDC/FIS-AST/4862/2020 and work contract 2020.03379.CEECIND.

The research here made use of the
IDL\footnote{\url{https://www.l3harrisgeospatial.com/Software-Technology/IDL}},
Python\footnote{\url{https://python.org}}, in particular
\texttt{numpy} \citep{harris2020array} and
Astropy,\footnote{\url{http://www.astropy.org}} a community-developed core
Python package for Astronomy \citep{astropy:2013, astropy:2018}, 
Perl\footnote{\url{https://www.perl.org}},  and
Julia\footnote{\url{https://julialang.org/}} \citep{Julia-2017} programming
languages. I also made use of the DS9
\citep{joyeNewFeaturesSAOImage2003} image visualization program.

\section*{Data Availability}

The analysis presented here is based on the publicly available L3 data
products provided in the Barbara A. Mikulski Archive for Space
Telescopes (MAST), as well as datasets provided in the literature.

Tables with line fluxes are available from
\url{https://github.com/jbrinchmann/JWST_SMACS}. The \texttt{PIFit}
code is currently in pre-beta and is available upon reasonable request
but will be publicly released in the near future. The \texttt{CL01Fit}
and \texttt{platefit} IDL codes are available upon reasonable request,
but a Python version of \texttt{platefit} will released with Bacon et
al (submitted).



\bibliographystyle{mnras}
\bibliography{smacs-ero} 

\bsp	
\label{lastpage}
\end{document}